\documentstyle[11pt,preprint,prd,aps]{revtex}

\tighten

\begin{document}

\draft

\title{Non-universal GUT corrections to
the soft terms and their implications in supergravity models}

\author{Nir Polonsky and Alex Pomarol\thanks{
Address after September 1 1994: Theory Division, CERN, CH-1211
Geneva 23, Switzerland.}
}
\address{Department of Physics, University of Pennsylvania,
Philadelphia, Pennsylvania, 19104, USA}
\date{September 1994, UPR-0627T}

\maketitle

\begin{abstract}
Potentially large
non-universal corrections to the soft supersymmetry breaking
parameters arise from their evolution
between the Planck and the grand-unification scales.
We detail typical patterns of non-universality in GUT models,
as well as elaborate on their propagation to the weak scale
and on their low-energy implications.
Possible corrections to the different scalar quark and lepton
masses and the Higgs and the gaugino-Higgsino sector parameters
are described in detail, and new allowed regions
of the parameter space are pointed out.
In particular, the patterns studied
often lead to heavier Higgsinos and $t$-scalar.
One-loop GUT threshold corrections to the soft parameters
are also discussed and shown to be important.
\end{abstract}


\section{Introduction}
\label{sec:s1}

It was recently pointed out  \cite{us}
that in minimal supergravity type models
\cite {review}, the model-dependent  renormalization of the soft
mass parameters between the Planck scale
$M_{P} \approx 2 \times 10^{18}$ GeV
and the grand-unification scale $M_{G} \approx  2 \times 10^{16}$  GeV,
can significantly modify the boundary conditions
for these parameters at $M_{G}$.
In particular, contrary to the standard working assumption
of universality at $M_{G}$
($e.g.$, see Ref. \cite{examples,penn594}),
specific patterns of non-universality are induced
at $M_{G}$ even when the soft parameters have universal
boundary conditions at $M_{P}$.
Hence, low-energy predictions and constraints in
the minimal supersymmetric extension of the standard model (MSSM)
\cite{review} are subject to model-dependent modifications.
Regions of the parameter space that are of interest
to present and future collider experiments
may change and/or be smeared, requiring one to assign
uncertainties to the MSSM low-energy predictions.
On the other hand, the discovery of superpartner and Higgs
particles could provide new and exciting hints on the physics
near the Planck scale.

In general, as a result of the GUT effects
the soft masses
can be different at $M_{G}$
for fields which are in different representations of the
unified gauge group. This is due to $(i)$ the different charges
and $(ii)$ the different Yukawa
interactions of the different fields.
Indeed, if one assumes that the corrections are
proportional to only $(\alpha_{G}/\pi)\ln(M_{P}/M_{G}) \sim 5\%$,
then the effects are negligible. However,
the above argument does not hold, regardless of the size of the
Yukawa interactions, if large representations
are present. In such a case,  the unified coupling
$\alpha_{G}$ is multiplied by a large number and
the effect of Planck to GUT scale evolution can be significant
\cite{earlywork,us}.
Furthermore, in Ref. \cite{us} it is shown that top and bottom
Yukawa couplings, as well as GUT-scale Yukawa couplings, which
have to be large to avoid a too rapid proton decay \cite{proton},
can also induce large deviations from universality at $M_{G}$.
In particular, the soft supersymmetry
breaking (SSB) parameters related to
the light Higgs fields can be significantly different
from those related to the
matter fields due to the different Yukawa interactions.
(It should be emphasized that universality of the soft mass parameters
was assumed, but at $M_{P}$.)
Moreover, once non-universality exists at $M_{G}$,
and if the rank of the gauge group
is higher than that of the standard model (SM) group, then non-vanishing
$D$-terms \cite{dterms} exist \cite{newdterms}.
The $D$-terms, which are charge dependent, induce a secondary breaking
of universality.

Given the above, one may then question the motivation to assume universality
at any scale: Relaxation of that assumption is subject to strong constraints
from flavor changing neutral currents (FCNC) \cite{fcnc}.
Also, any predictive power
is lost. On the other hand, the identification of
the universality scale
with $M_{G}$ is convenient but ad hoc,
and the consequences of its relaxation need exploration.
Allowing $M_{P} - M_{G}$ renormalization of the parameters
leads to restricted patterns of non-universality at $M_{G}$.
In particular, the super-GIM mechanism suppressing FCNC
in the MSSM \cite{su5s}
need not be altered if Yukawa couplings of the first and
second families remain negligible at all scales.
In addition, the predictive power is
altered but not lost. In SU(5) models,
one has to introduce at least two more
Yukawa couplings, and for higher-rank groups, $e.g.$, SO(10),
a new parameter is needed to account for the magnitude and sign
of the $D$-terms.
The deviations from universality at $M_{G}$
are not arbitrary but are calculable in terms of the
new parameters.
However, the interference between the different corrections, $e.g.$,
those from Yukawa interactions and those from $D$-terms
(and in particular, if higher order gravitational corrections
are not negligible),
can render it difficult to disentangle traces of the high-scale theory
in the low-energy physics.

Below, we will again resort to the assumption of universality
of the SSB parameters at\footnote{
That choice maximizes the effects.
Universality at a lower scale
would lead to lesser but similar effects.
We do not consider gravitational effects other than those which induce
the universal soft terms. Higher order gravitational
effects could add to the uncertainty.}
$M_{P}$ (the obvious choice in minimal
supergravity theories \cite{old}),
$i.e.$, a common scalar mass $m_{0}$, a common gaugino mass
$M_{1/2}$, and common dimension-one trilinear and bilinear scalar couplings
$A_{0}$ and $B_{0}$.
(In fact, our discussion is independent of any assumptions
regarding universality of the bilinear couplings $B_{i}$.)
We will then assume a grand unified theory (GUT)
between $M_{P}$
and $M_{G}$. ($M_{G}$
is determined by gauge coupling unification.)
Note that
the universality of the gaugino masses above $M_{G}$ is trivial
if the GUT group is simple.
For simplicity, we assume in most of our calculations the minimal SU(5)
model \cite{su5,su5s}.
However, extended models, including models with
non-vanishing $D$-terms, are studied qualitatively.
The effective theory below $M_{G}$ is assumed to be the MSSM
with the appropriate $M_{G}$ matching conditions.
We then study the different patterns of non-universality which
are induced at $M_{G}$, their propagation to the weak scale,
low-energy implications and consequences for model building.
In particular, we will examine two points in the parameter space,
\begin{flushleft}
$(a)$ $m_{t}^{pole} = 160$ GeV, $\tan\beta = 1.25$;
\newline
$(b)$ $m_{t}^{pole} = 180$ GeV, $\tan\beta = 42$;
\end{flushleft}
for the $t$-quark pole mass $m_{t}^{pole}$ and for $\tan\beta = \langle H_{2}
\rangle / \langle H_{1} \rangle$.
We choose these points because of the large
$t$ and $b$ quark Yukawa couplings, $i.e.$,
$h_{t} \approx 1 \gg h_{b}$ and $1 \gtrsim h_{t} \gtrsim h_{b}$,
respectively.
Also, because of the large Yukawa couplings,
points $(a)$ and $(b)$ are consistent with
bottom-tau unification and with minimal SU(5), $e.g.$, see Ref.\ \cite{us556}.
For intermediate values of $\tan\beta$ the effects are
a superposition of those for points $(a)$ and $(b)$.

It was recently suggested that non-universality of the soft terms
is a typical signature of some string models \cite{string1}.
In string models one often has
a direct unification at the string scale $M_{S} < M_{P}$,
$i.e.$,
$M_{G} = M_{S}$, and the universal or non-universal
boundary conditions are derived from the string
theory
at $M_{S}$.
However, the string scale is typically
$M_{S} \approx 5.2 \times g_{S} \times 10^{17}$ GeV \cite{mstring},
and the relation $M_{G} = M_{S}$ fails for the MSSM.
String inspired non-universality is studied in
Ref.\ \cite{string2}. In particular, Kobayashi et al.
suggest a solution to the
$M_{S}/M_{G} \approx 20$ discrepancy.
If one assumes that the string theory leads to an intermediate
GUT, then our analysis applies, but the results should be scaled
down by $\sim \ln{(M_{S}/M_{P})}$.

We previously found \cite{us} that some low-energy parameters such as
the $\mu$ parameter and the $t$-scalar mass, can be significantly
modified while others, $e.g.$, the SM-like Higgs boson mass, are nearly
invariant under minimal SU(5) type corrections.
We also pointed out the $\tan\beta$-dependent modification of the allowed
parameter space and of correlations between different observables.
Here, we will further elaborate on the above observations and expand
our previous work.
We review the possible patterns of non-universality
and compare their generation via radiative corrections
to the radiative symmetry breaking mechanism
in section \ref{sec:s2}.
We also demonstrate that the same rule of thumb
holds for the $M_{P} - M_{G}$ and $M_{G} - M_{Z}$
evolutions of the soft parameters and their resulting
hierarchy: It is determined by the competition
between the gauge charge and Yukawa interactions
of the relevant field and by the asymptotic freedom of the model.
In section \ref{sec:s3} we describe our numerical routines and
discuss the weak-scale phenomena.
In particular, we elaborate on the
propagation of non-universal corrections from
$M_{G}$ to the weak scale; on the
distinction between physical (bottom-up approach)
and model-building (up-down approach) parameters;
and on the implications
to the first and second family scalars,
the $\mu$ parameter,
the Higgs sector, and third family scalars.
Future observations of the signatures described could support the
existence of an intermediate GUT. Their absence, on the other hand,
could indicate direct unification,
but also interference between
different effects or small values for the new parameters.
Our conclusions are given in Section
\ref{sec:s4}. For completeness, the minimal SU(5)
model is defined in Appendix \ref{sec:a1}.
Threshold corrections (due to the
ambiguity in $M_{G}$) in that model are described and discussed
in Appendix \ref{sec:a2}.

Rather than restrict oneself to the patterns described in section \ref{sec:s2},
one could adopt a more phenomenological approach to non-universality
by postulating certain universality breaking patterns.
For example, Dimopoulos and Georgi considered different boundary conditions
for the matter and Higgs bosons \cite{su5s}. Similar approaches
were adopted recently by several authors.
In Ref.\ \cite{o1} a split
at $M_{G}$ between the light Higgs and matter fields is considered
[but mainly in the context of SO(10) scenarios].
In Ref.\ \cite{o2} patterns of non-universality
desired in certain SO(10) models are studied.
A general discussion of non-universal
scalar potentials at $M_{G}$ was recently given in Ref.\ \cite{o3}.
Other recent studies of non-universality were carried
out in Ref.\ \cite{o4}.
Where relevant, the conclusions of these authors agree with ours.


\section{Patterns of non-universality at $M_{G}$}
\label{sec:s2}

Before discussing the way in which the Planck to GUT scale
evolution of the soft mass parameters can induce large
deviations from universality at $M_{G}$, it is useful to recall
the more familiar
way in which GUT to weak scale evolution can render
the weak-scale scalar potential consistent with
a spontaneously broken SU(2)$_{L}$$\times$U(1)$_{Y}$ symmetry \cite{rsb}
[often called the radiative symmetry breaking (RSB) mechanism].
In both cases large Yukawa couplings play a similar role
and lead to similar behavior of the SSB parameters.
The RSB mechanism is easily understood if one writes the approximate
renormalization group equations (RGEs)
of the soft scalar masses \cite{rge}:
\begin{equation}
\frac{d}{d\ln{\cal Q}}
\left[ \begin{array}{c}
m^{2}_{H_{1}} \\
m^{2}_{H_{2}} \\
m^{2}_{U} \\
m^{2}_{Q}
\end{array} \right]
=
\frac{h_{t}^{2}}{8\pi^{2}}
\left[ \begin{array}{cccc}
0&0&0&0\\
0&3&3&3\\
0&2&2&2\\
0&1&1&1
\end{array} \right]
\left[ \begin{array}{c}
m^{2}_{H_{1}} \\
m^{2}_{H_{2}} \\
m^{2}_{U} \\
m^{2}_{Q}
\end{array} \right]
- \frac{2g_{3}^{2}}{3\pi^{2}}M_{\tilde{g}}^{2}
\left[ \begin{array}{c}
0\\
0\\
1\\
1
\end{array} \right],
\label{rge1}
\end{equation}
where  ${\cal Q}$ is the renormalization scale,
$g_{3}$ is the SU(3)$_{c}$ coupling and
\begin{equation}
- {\cal L}_{soft} = \sum_{i} m_{i}^{2}|\phi_{i}|^{2}
+[B\mu H_{1}H_{2}
+ A_{t}h_{t}QH_{2}U +
\frac{1}{2}\sum_{\lambda}M_{\lambda}\lambda_{\alpha}\lambda_{\alpha}
\, + h.c. ],
\label{lsoft}
\end{equation}
with $i$ ($\lambda$) summing over all scalars (gauginos)
and $H_{1}$, $H_{2}$, $ U = \tilde{t}_{R}$,
$Q = (\tilde{t}_{L}, \, \tilde{b}_{L})$,
and $\tilde{g}$
are the down and up type Higgs doublets, right-handed
$t$-scalar, left-handed scalar-quark doublet
and the gluino, respectively.
(Below, we will also refer to the $t$-scalar, scalar quarks, etc. as
stop, squarks, etc.)
$\mu$ is the mass parameter in the MSSM superpotential,
eq.\ (\ref{effsuper}).
For simplicity we omitted in (\ref{rge1}) $A_{t}$-term contributions and
neglected all terms aside from the
QCD and $h_{t}$ Yukawa terms which typically dominate the evolution.
(As we discuss below, in some cases the $h_{b}$ term can be equivalent
to the $h_{t}$ term and has to be considered as well.)
The stop masses grow with the gluino mass
$M_{\tilde{g}}({\cal Q}_{1}) =
[\alpha_{3}({\cal Q}_{1})/\alpha_{3}({\cal Q}_{2})]
M_{\tilde{g}}({\cal Q}_{2})$,
a growth which is subject to some slow-down due to the Yukawa term.
On the other hand, the Yukawa term diminishes $m_{H_{2}}^{2}$
to negative values
at the weak scale such that the sum
$m_{H_{2}}^{2} + \mu^{2} \sim {\cal O}(M_{Z}^{2})$.
($m_{H_{1}}^{2}$
is not renormalized in this approximation.)
The global minimum of the weak-scale
Higgs potential is consistent with electroweak symmetry breaking (EWSB)
provided that
$(m_{H_{1}}^{2} + \mu)(m_{H_{2}}^{2} + \mu^{2}) \leq B^{2}\mu^{2}$
(and that
$m_{H_{1}}^{2} + m_{H_{2}}^{2} + 2\mu^{2} \geq 2 |B\mu|$),
which is indeed the situation
for ${\cal O}(M_{Z}^{2})$ (possibly negative) values of
$m_{H_{2}}^{2} + \mu^{2}$. Also, there are no
tachions in the theory.
The situation is, of course, more complicated when including all terms,
but is qualitatively similar to the above approximation.

For future reference, observe that (\ref{rge1}) is independent
of either $\mu$ or $B$. The decoupling of $\mu$ from the
SSB parameters and of $B$ from all other SSB parameters holds in general.
This enables one to rewrite the EWSB minimization conditions \cite{rsb} as
\begin{mathletters}
\label{min}
\begin{equation}
\mu^{2} = \frac{m_{H_{1}}^{2}
- m_{H_{2}}^{2}\tan^{2}\beta}{\tan^{2}\beta - 1}
-\frac{1}{2}M_{Z}^{2},
\label{min1}
\end{equation}
\begin{equation}
B\mu = -\frac{1}{2}\sin2\beta\left[
{m_{H_{1}}^{2} + m_{H_{2}}^{2} + 2\mu^{2}}\right].
\label{min2}
\end{equation}
\end{mathletters}
Hence, $B_{0}$ can be traded for $\tan\beta$ and $\mu$
is predicted as a function of the SSB parameters and of $\beta$.
(Triviality limits give $\tan\beta > 1$ for $m_{t}^{pole}
\gtrsim 140$ GeV.)

To summarize,
the gauge term dominates the $M_{G} - M_{Z}$ evolution
for the colored scalar soft masses (but their spectrum
also
bears traces of their Yukawa interactions) while the Yukawa
term dictates $m_{H_{2}}^{2}$ evolution
(in practice, it dominates over  the $H_{2}$ weak and hypercharge
gauge terms). Thus, a large hierarchy is generated at the weak scale,
even when assuming universal scalar masses
as the GUT scale boundary condition for (\ref{rge1}).
Let us now write in a similar fashion the RGEs in the minimal
SU(5) model, which is assumed to dictate the $M_{P} - M_{G}$
evolution
[the complete RGEs in that model
are given in \cite{us} and in Appendix \ref{sec:a1}]:
\begin{equation}
\frac{d}{d\ln{\cal Q}}
\left[ \begin{array}{c}
m^{2}_{{\cal H}_{1}} \\
m^{2}_{{\cal H}_{2}} \\
m^{2}_{10}
\end{array} \right]
=
\frac{h_{t}^{2}}{8\pi^{2}}
\left[ \begin{array}{cccc}
0&0&0\\
0&3&6\\
0&3&6
\end{array} \right]
\left[ \begin{array}{c}
m^{2}_{{\cal H}_{1}} \\
m^{2}_{{\cal H}_{2}} \\
m^{2}_{10}
\end{array} \right]
+
\frac{3\lambda^{2}}{5\pi^{2}}
\left[ \begin{array}{cccc}
1&1&0\\
1&1&0\\
0&0&0
\end{array} \right]
\left[ \begin{array}{c}
m^{2}_{{\cal H}_{1}} \\
m^{2}_{{\cal H}_{2}} \\
m^{2}_{10}
\end{array} \right]
- \frac{g_{G}^{2}}{5\pi^{2}}M_{5}^{2}
\left[ \begin{array}{c}
6\\
6\\
9
\end{array} \right],
\label{rge2}
\end{equation}
where $m_{{\cal H}_{1}}$ and $m_{{\cal H}_{2}}$ are the soft masses of the
$\bar{\bf 5}$ and {\bf 5} SU(5) Higgs bosons
(that contain $H_{1}$ and $H_{2}$, respectively); $m_{10}$ and $m_{5}$
(that we use later) are the soft masses of the
{\bf 10} (that contains $Q$ and $U$) and $\bar{\bf 5}$ matter superfields;
$g_{G}$ and $M_{5}$ are the SU(5) gauge coupling and gaugino mass,
and the Yukawa coupling $\lambda$ is defined in Appendix \ref{sec:a1}.
Note that the introduction of larger representations typically implies
larger numerical coefficients.
The GUT effects in the minimal SU(5) model
can be read from (\ref{rge2}) and
are described by the following patterns:
If $h_{t} \approx \lambda \approx 1$ at $M_{G}$
(which is consistent with these couplings being
$\lesssim 2$ near $M_{P}$, so the one-loop approximation
is reasonable) then the $\lambda$ term diminishes the ${\cal H}_{1}$
and ${\cal H}_{2}$ squared soft masses significantly
while the $h_{t}$ term lifts their degeneracy,
further diminishing $m_{{\cal H}_{2}}^{2}$.
The latter also diminishes $m_{10}^{2}$.
In comparison to (\ref{rge1}), the additional Yukawa term,
the larger numerical coefficients
and $\alpha_{G} \ll \alpha_{3}(M_{Z})$
compensate for the shorter evolution interval,
leading to a similar pattern (aside from the effect
on ${\cal H}_{1}$), but at $M_{G}$  rather than at $M_{Z}$,
and we observe the pattern
\begin{flushleft}
$(i)$  $m_{5}^{2} \geq m_{10}^{2} \gg
m_{{\cal H}_{1}}^{2} \geq m_{{\cal H}_{2}}^{2}$.
\end{flushleft}
First and second family scalar masses are
renormalized to a very good approximation only by
gauge interactions, and hence
are slightly heavier
than the third family $m_{5}^{2}$ (which is renormalized,
in practice, by the $h_{b}$ term).
More importantly, in this scenario
the gauge interactions  do not lift the
degeneracies between the first and second family SSB parameters
and the MSSM super-GIM mechanism is still intact.
Pattern $(i)$ would now serve as a non-universal boundary condition
to (\ref{rge1}). As can be seen in Fig.\ 1a of Ref. \cite{us}
and in Fig.\ \ref{fig:fig1}  here,
$m_{{\cal H}_{2}}^{2}$ can be driven to near zero values
already at $M_{G}$, and in principle, RSB could be achieved
for small values of $h_{t}$ (this is, of course, irrelevant
for $m_{t}^{pole} \gtrsim 100$ GeV).
Before proceeding, let us stress that
the hierarchy between the different
SSB parameters indeed depends on
the gauge charges and on the size of the Yukawa couplings.
However, whether the parameters grow or diminish depends
roughly on the ratio $[g_{G}^{2}M_{1/2}^{2}]/
[\max(h_{t}^{2},\,\lambda^{2})\times m_{0}^{2}]$.
If the ratio is larger than unity, then typically
all the parameters grow with decreasing energy.
This is true in general and is seen
[for pattern $(i)$] in Fig.\ 1b of Ref.
\cite{us} [where
$g_{G}^{2}M_{1/2}^{2} \gg h_{t}^{2}m_{0}^{2}
\Rightarrow m_{10}(M_{G}) > m_{5}(M_{G})$]
and in Fig.\ \ref{fig:fig2}, here. (This is always the case
for the first and second family scalars.)
If both, $h_{t}$ and $\lambda$, are large, as in pattern $(i)$,
then for small and moderate values of $M_{1/2}$
only $m_{5}^{2}$ grows.
(Fig.\ \ref{fig:fig2} corresponds
to the no-scale assumption $m_{0} = A_{0} = 0$.)

The GUT effects leading to pattern $(i)$ are a mismatch
of effects due to the large $\lambda$ and to the large $h_{t}$.
Different assumptions regarding the Yukawa couplings lead
to different patterns.
If $h_{t}(M_{G}) \ll 1$ ($i.e.$, $m_{t}^{pole} \lesssim 180$ GeV
and $\tan\beta \gg 1$) and also $h_{b}(M_{G})$
is small ($i.e.$, $\tan\beta
\lesssim 40$), then a simpler pattern arises,
\begin{flushleft}
$(ii)$  $m_{10}^{2}, \, m_{5}^{2} >
m_{{\cal H}_{1}}^{2}, \, m_{{\cal H}_{2}}^{2}$.
\end{flushleft}
The splitting between the Higgs and matter sectors
depends, as before, on the size of $\lambda$.
(We comment on the case $h_{b} > h_{t}$ below.)
Regarding $\lambda$, in the minimal SU(5) model one has
$\lambda = g_{G}M_{H_{C}}/M_{V}$ (see Appendix A).
In this model, proton decay non-observation requires
that the colored triplet is heavy $M_{H_{C}} \gtrsim
M_{V}$ \cite{proton} and hence $\lambda \approx 1$.
Thus, patterns $(i)$ and $(ii)$ completely
characterize that model.
However, if we ignore the proton decay constraint,
assume $\lambda \ll 1$
and again take $h_{t} \gg h_{b}$, we find
($e.g.$, see Fig.\ \ref{fig:fig3})
\begin{flushleft}
$(iii)$  $m_{5}^{2},\,
m_{{\cal H}_{1}}^{2} >
m_{10}^{2},\,
m_{{\cal H}_{2}}^{2}$.
\end{flushleft}
Finally, if all Yukawa couplings are small
one has
\begin{flushleft}
$(iv)$  $m_{10}^{2} >
m_{5}^{2},\,
m_{{\cal H}_{1}}^{2},\,
m_{{\cal H}_{2}}^{2}$.
\end{flushleft}

It is convenient to relate the different patterns
(or GUT effects) to values of $\tan\beta$. Patterns $(i)$
and $(iii)$ correspond to low $\tan\beta \approx 1 - 2$,
$i.e.$, choice $(a)$.
Patterns $(ii)$ and $(iv)$  correspond to moderate
values of $\tan\beta$.
An interesting scenario arises in the special case
$h_{t} \approx h_{b} \approx 1$ (and $\lambda \approx 1$),
$i.e.$, large $m_{t}^{pole} \gtrsim 180$ GeV and
large $\tan\beta \approx 50 - 60$.
Eq.\ (\ref{rge2}) now has an additional Yukawa term
previously neglected: The $h_{b}$ term
(see Appendix \ref{sec:a1}).
This again leads to pattern $(ii)$, however, the
corrections are now enhanced
because the three Yukawa couplings,
$\lambda$, $h_{t}$ and $h_{b}$, are large, and unless one assumes
a small $\lambda$ and
a specific mechanism to suppress proton decay
(or $M_{1/2} \gg m_{0}/g_{G}$),
one has a special case of
$(ii)$, $i.e.$,  $m_{0}^{2} \gg m_{5}^{2}, \, m_{10}^{2} >
m_{{\cal H}_{1}}^{2} \sim m_{{\cal H}_{2}}^{2}$.
That situation is similar to the one in the minimal
SO(10) model where the Higgs {\bf 5} + $\bar{\bf 5}$ of SU(5)
are embedded in a single {\bf 10} of SO(10)
(and {\bf 10} + $\bar{\bf 5}$ in a  {\bf 16}).
Yukawa unification requires in that case
$h_{t} \approx h_{b} \sim 1$.
Our choice $(b)$ with $h_{t}(M_{G}) \sim 0.5$ and $h_{b}(M_{G}) \sim 0.3$
corresponds to a moderate version of this scenario.
A scenario corresponding to $(b)$
is illustrated in Fig.\ \ref{fig:fig4}.
Of course, in SO(10) the RGEs have slightly different slopes
than in our case, and the magnitude of the splittings
calculated in SU(5) can only approximate the actual splittings
in SO(10).
The different patterns and their dependence on $\tan\beta$
are summarized in Table \ref{table:t0}.

We now turn to examine the situation in some extended GUT models.
When considering extended models one could
assume a higher rank group, additional (and/or larger) representations,
or both.
As an example of the latter,
in non-minimal SU(5) (and other) models
${\cal H}_{1}$ and ${\cal H}_{2}$ couple with different strength to
the other Higgs superfields
[below we refer to this scenario as non-minimal SU(5)].
One could now arrange for a large and  arbitrary splitting at $M_{G}$
between $m_{{\cal H}_{1}}^{2}$ and $m_{{\cal H}_{2}}^{2}$, $e.g.$,
\begin{flushleft}
$(v)$  $m_{5}^{2} > m_{10}^{2} >
m_{{\cal H}_{i}}^{2} \gg m_{{\cal H}_{j}}^{2}$,
\end{flushleft}
where $i \neq j = 1,\,2$.
This can provide a caveat for the general rule that
no RSB is possible (assuming universality)
for $h_{b} > h_{t}$. In fact, we confirmed that the splitting
in $(v)$ can be arranged (for $i = 1$)
so that RSB is possible in that case.
For example,
in the SU(5) missing partner model (MPM) \cite{mpm1},
the superpotential reads
$W = \lambda_{1}{{\cal H}_{1}}\Sigma({\bf 75})\Phi({\bf 50})
+  \lambda_{2}{{\cal H}_{2}}\Sigma({\bf 75})\Phi(\bar{\bf 50}) + $...
Proton decay constraints in that model \cite{mpm2}
can be approximated as $\lambda_{1}\lambda_{2} \gtrsim 5\eta g_{G}$
with $\eta \sim M_{\Phi} / \langle 75 \rangle$
in the range $0.1 \lesssim \eta \lesssim 10$.
However, some caution is in order.
One has to keep in mind a possible breakdown of
perturbation theory when large representations
(or a large number of small representations) are present.
For example, the MPM is not asymptotically free and
the gauge coupling typically diverges below $M_{P}$.
(Texture models often assume large representations
when trying to explain the light fermion spectrum and
may not be asymptotically free as well.)
In general, one needs to develop
non-perturbative techniques in order to calculate
the GUT effects to the SSB parameters.
In non-asymptotically free but still perturbatively
valid models one has $M_{5}(M_{G}) \ll M_{1/2}$.

Before turning to discuss higher rank groups,
the $M_{G}$ matching conditions
for SU(5) (minimal and non-minimal) models
read
\begin{mathletters}
\label{su5}
\begin{equation}
m^{2}_{Q} =
m^{2}_{U} =
m^{2}_{E} =
m^{2}_{{10}},
\label{su5a}
\end{equation}
\begin{equation}
m^{2}_{D} =
m^{2}_{L} =
m^{2}_{{5}},
\label{su5b}
\end{equation}
\begin{equation}
m^{2}_{H_{1,\,2}} =
m^{2}_{{\cal H}_{1,\,2}},
\label{su5c}
\end{equation}
\end{mathletters}
where $Q = (\tilde{t}_{L}, \,\tilde{b}_{L})$
and $L = (\tilde{\nu_{\tau}}_{L}, \,\tilde{\tau}_{L})$
are the left-handed scalar quark and lepton doublets,
and $U$, $D$ and $E$ are the right-handed
$t,\,b,$ and $\tau$-scalars, respectively.
Similar relations hold  for the first and second families.
As was shown above,
in the minimal model, $m^{2}_{{\cal H}_{1,\,2}}$
both depend on $\lambda$
and are correlated. In non-minimal models, $e.g.$,
the MPM, they are independent parameters.
Condition (\ref{su5}) applies to
the RG-improved tree level SSB parameters\footnote{
The tree-level SSB parameters shift
the GUT-scale vacuum expectation values and  redefine $\mu$, $B$
(recall that we are not required to specify these parameters
at the high scale),
as well as induce the D-terms \cite{mu,dterms,newdterms}.}
at $M_{G}$.
However, since not all heavy fields are degenerate at $M_{G}$,
the splitting between the heavy masses induces, at one-loop, a secondary
non-universal
shift in the soft parameters,
{\bf regardless} of their initial universality \cite{us}.
This is similar to threshold corrections
to dimensionless couplings, $e.g.$, see Ref.\ \cite{us513,us556,mpm2},
with the exception
that corrections to $m_{i}^{2}$ due to boson-fermion mass splittings
are now $\sim M_{G}^{2}\ln[(M_{G}^{2} + m_{soft}^{2})/M_{G}^{2}]
\sim m_{soft}^{2}$ and are not suppressed.
One-loop threshold corrections could smear the above patterns,
and are discussed for the minimal SU(5) model in Appendix
\ref{sec:a2}.

If the rank of the GUT group
is higher than that of the SM group  [or for that matter, of SU(5)],
$i.e.$, higher than four, non-universality at $M_{G}$
would, in general, trigger non-vanishing $D$-terms \cite{dterms,newdterms}
that could correct the $M_{G}$ boundary conditions.
(The $D$-terms appear at the scale
at which the rank of the group is reduced.)
It is important to stress that had we assumed universality
at $M_{G}$ then the $D$-terms would have vanished
leaving the universality assumption intact \cite{newdterms}.
We consider, as an example,
the minimal SO(10) model
which now depends on an additional parameter --
the magnitude (and sign) of the $D$-terms $M_{D}^{2}$,
which can be shown to be of the order of the soft mass parameters.
In SO(10), condition (\ref{su5})
is replaced by \cite{dterms,newdterms}
\begin{mathletters}
\label{so10}
\begin{equation}
m^{2}_{Q} =
m^{2}_{U} =
m^{2}_{E} =
m^{2}_{{16}} + M_{D}^{2} \equiv m_{10}^{2},
\end{equation}
\begin{equation}
m^{2}_{D} =
m^{2}_{L} =
m^{2}_{{16}} - 3M_{D}^{2} \equiv m_{5}^{2},
\end{equation}
\begin{equation}
m^{2}_{H_{1,\,2}} =
m^{2}_{{\cal H}_{1,2}} \pm 2M_{D}^{2},
\end{equation}
\end{mathletters}
and in the minimal\footnote{Note, however, that in some texture models
Higgs doublets are embedded, $e.g.$, in {\bf 10}'s and in {\bf 126}'s,
and $m^{2}_{{\cal H}_{1}}$  and $m^{2}_{{\cal H}_{2}}$ could
evolve very differently.}
SO(10) model
$m^{2}_{{\cal H}_{1}} = m^{2}_{{\cal H}_{2}} \equiv  m^{2}_{\cal H}$
and typically $m_{0}^{2} \gg m_{16}^{2} \gg m^{2}_{\cal H}$
(recall the very large $\tan\beta$ case).
$m_{H_{1}}^{2}$ and $m_{H_{2}}^{2}$
could be split at $M_{G}$
according to the sign of $M_{D}^{2}$.
[Note the SU(5) invariance of (\ref{so10}).]
A situation similar to (\ref{so10}) could arise in a
SU(5)$\times$U(1) model.

Below, we discuss the minimal SU(5) model [patterns $(i) - (iv)$]
as well as
non-minimal SU(5) and
minimal SO(10)-inspired
extensions [pattern $(v)$ and eq.\ (\ref{so10}), respectively].
Our choices of points $(a)$ and $(b)$ correspond
(for $\lambda \approx 1$) to patterns $(i)$
and $(ii)$, respectively.
Hereafter, that correspondence is understood when discussing
low and large values of $\tan\beta$.
The latter can be used as a crude approximation
of the situation in the minimal SO(10) model.
Intermediate values of
$\tan\beta$ ($h_{t}$ and $h_{b}$ are both small)
are also described by patterns $(ii)$ [and $(iv)$]
and typically the splittings are somewhat diminished.


\section{Weak-scale phenomena}
\label{sec:s3}

In the previous section  we presented  four patterns of
$M_{G}$ boundary conditions, that define uniquely
non-universality in the minimal SU(5) model (up to threshold corrections).
To  analyze their low-energy implications one has to
evolve the SSB parameters from $M_G$ to $M_Z$.
The GUT corrections  to the scalar soft masses at $M_G$,
denoted by $\Delta m^2_i(M_G)$,
do not always lead to the same magnitude of corrections at $M_Z$.
For the scalar fields
${H_2}$, $Q$ and $U$  that couple
with a large Yukawa coupling ($h_t$),  the  GUT corrections
are usually diminished
in the evolution from $M_G$ to $M_Z$,
{\it i.e.},   $\Delta m^2_i(M_Z)< \Delta m^2_i(M_G)$.
For the rest of the fields, however, one has
$\Delta m^2_i(M_Z)\approx  \Delta m^2_i(M_G)$.

\begin{figure}
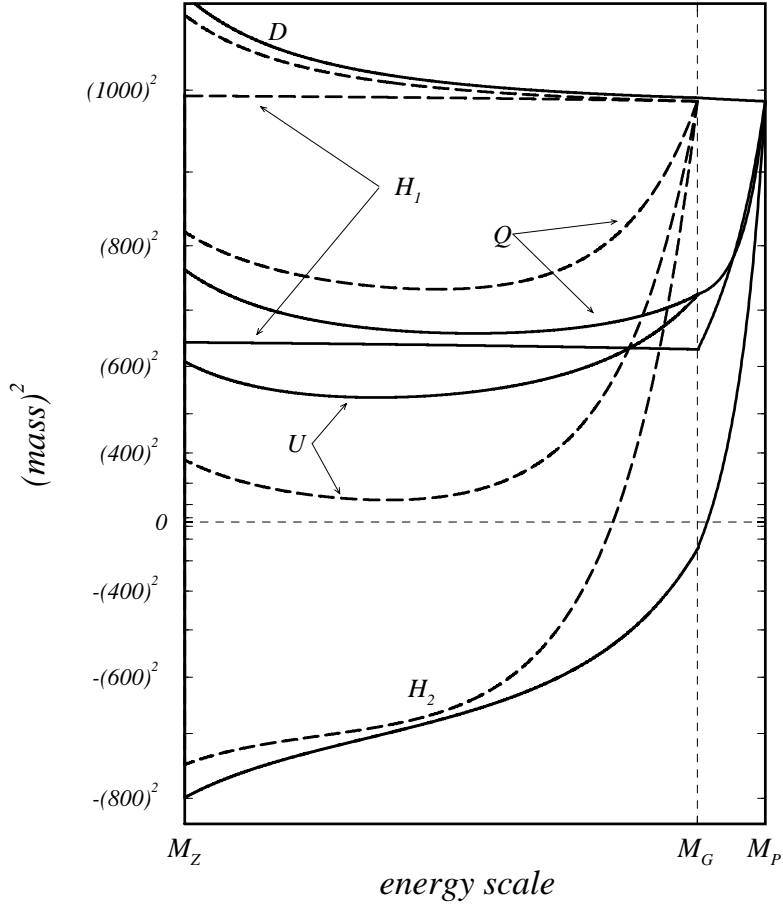

\caption{Evolution of the soft parameters $m_{i}^{2}$
corresponding to the two Higgs doublets $H_{1}$ and $H_{2}$
and third family scalar quark fields $Q$, $U$ and $D$
[assuming minimal SU(5)]
when $M_{P} - M_{G}$ evolution is
considered  (solid lines)
and neglected (dashed lines), plotted $vs.$
the logarithm of the energy scale.
The universal SSB parameters (taken at $M_{P}$ or $M_{G}$,
respectively) are
$M_{1/2} = 165$ GeV, $m_{0} = -\frac{1}{2}A_{0} = 987$ GeV,
and choice $(a)$:
$m_t^{pole} = 160$ GeV and $\tan\beta = 1.25$. Also, in the SU(5) case
we take $\lambda(M_{G}) = 1$ and $\lambda^{'}(M_{G}) = 0.1$.
}
\label{fig:fig1}
\end{figure}

\begin{figure}
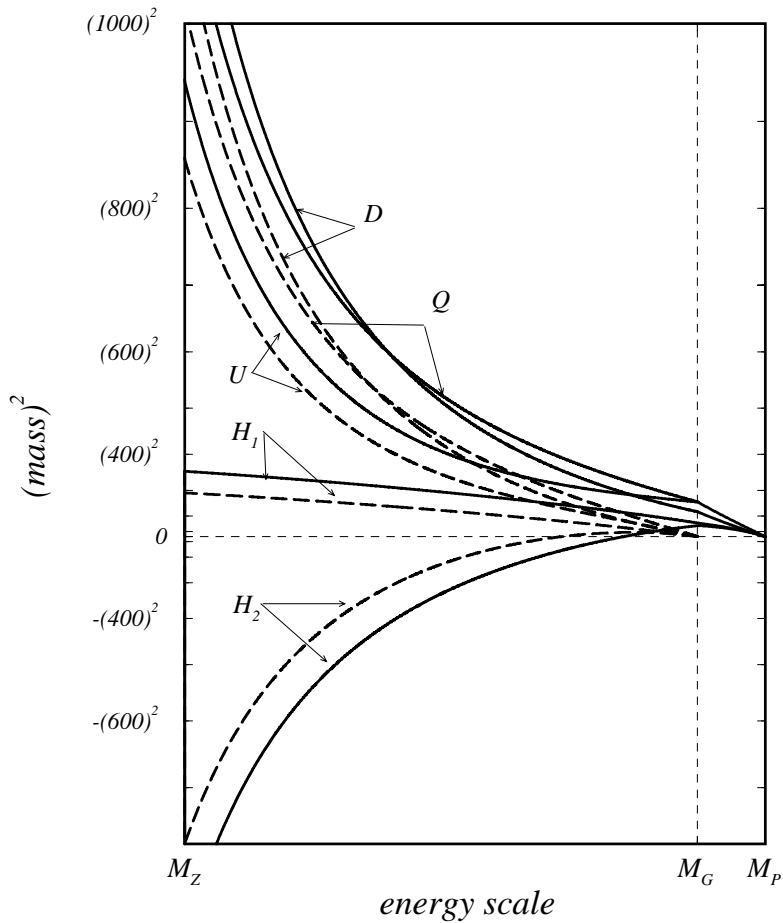

\caption{
Same as in Fig. 1 except
$M_{1/2} = 400$ GeV and  the no-scale assumption $m_{0} = A_{0} = 0$.
}
\label{fig:fig2}
\end{figure}

\begin{figure}
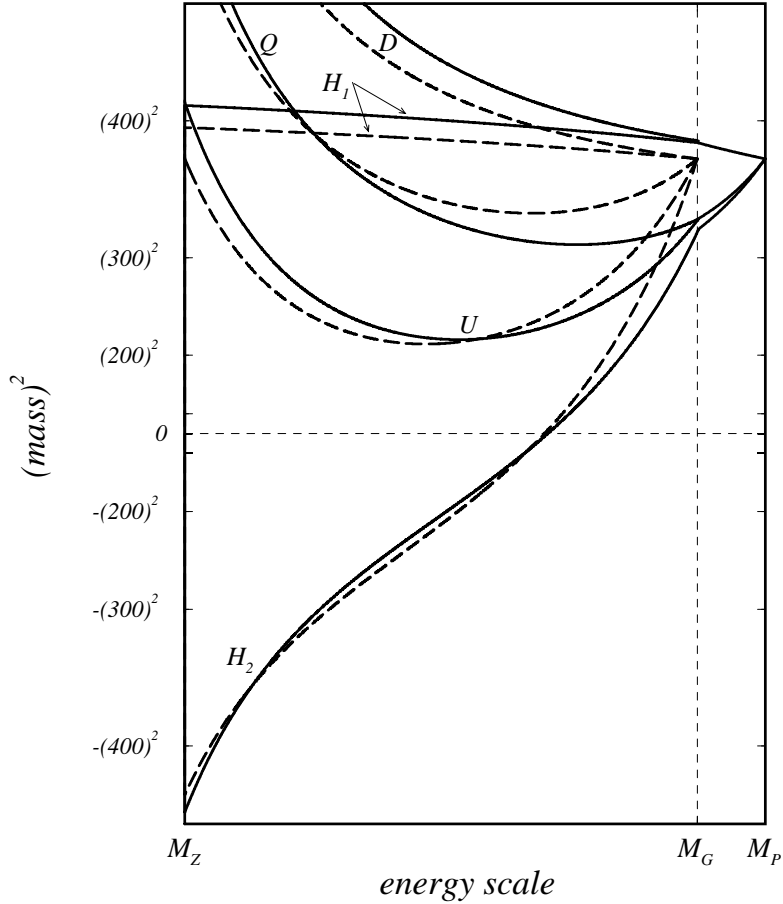

\caption{
Same as in Fig. 1 except
$M_{1/2} = 175$ GeV, $m_{0} = A_{0} = 375$ GeV and
$\lambda(M_{G}) = \lambda^{'}(M_{G}) = 0.1$.
}
\label{fig:fig3}
\end{figure}

\begin{figure}
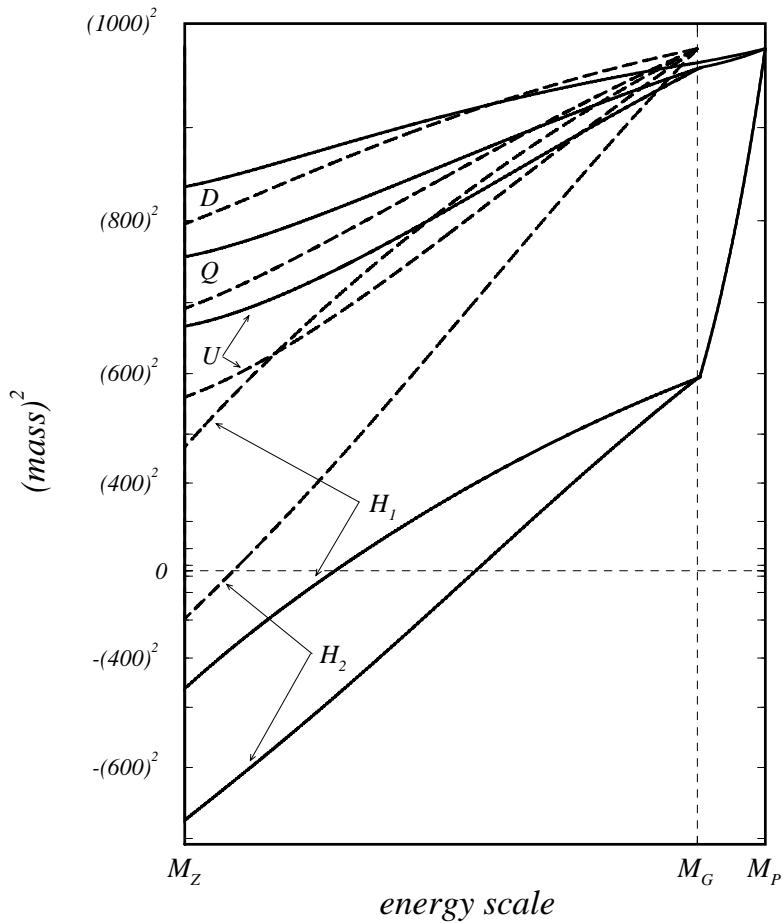

\caption{
Same as in Fig. 1 except
$M_{1/2} = 89$ GeV, $m_{0} = 977$ GeV, $A_{0} = 0 $,
and choice $(b)$: $m_t^{pole} = 180$ GeV and $\tan\beta = 42$.
}
\label{fig:fig4}
\end{figure}

Figs.\ \ref{fig:fig1} -- \ref{fig:fig4} show
the scale evolution of the soft scalar mass parameters.
The solid (dashed) line shows the evolution of the soft scalar masses
when the $M_P-M_G$ evolution is included (neglected).
Note that the GUT correction to $m^2_{H_2}$ (the difference between
the respective dashed and solid lines) is reduced at $M_Z$ when $h_t$ is large.
This is especially apparent in Figs.\ \ref{fig:fig1} and \ref{fig:fig3}.
In Fig.\ \ref{fig:fig2}, however,
the gauginos  give the largest contribution
to the  soft scalar masses and\footnote{Note that the gaugino mass
is enhanced $\sim 10\%$ by the $M_P-M_G$ evolution and that leads to
an additional increase in the $m^2_i(M_Z)$.
In Ref.\ \cite{o1,o2,o3,o4} only
the scalar soft masses are modified at $M_G$, $i.e.$,
$\Delta M^2_{5}(M_{G})=0$.}
$\Delta m^2_i(M_Z)-\Delta m^2_i(M_G)\propto\Delta M^2_{5}(M_{G})
\approx 20\%$.
This effect is especially important for colored particles because
of the gluino mass that grows at low energies with the
diverging QCD coupling and roughly triples between $M_{G}$
and $M_{\tilde{g}}$. This feeds back via gluino loops
to the colored scalar masses (and to the $A$ parameters)
 -- see (\ref{rge1}) --
leading to large renormalization of those SSB parameters at the weak scale.
Thus,  GUT corrections from the Yukawa sector [that can split fields
in equal SU(5) representations] can be washed out at $M_Z$.
When $h_t$ is smaller (Fig.\ \ref{fig:fig4}),
the reduction of $\Delta m^2_{H_2}$
is less drastic and  can give
rise to important low-energy implications (the same is true for
$\Delta m^2_{H_1}$ for any value of $\tan\beta$).
The evolution of  $\Delta m^2_{Q,\,U}$ is even more dramatic.
They can change the sign in the $M_G-M_Z$ evolution
and even (Fig.\ \ref{fig:fig4}) increase their value.
This is because the positive term
$\propto h^2_tm^2_{H_2}$ in the RGEs of  $m^2_{Q,\,U}$
(that decreases them with the energy scale)
is smaller when the GUT effects are considered.

Therefore, large effects  are expected
in those observables
that depend on $m^2_{H_i}$ and $m^2_{Q,\,U}$: The $\mu$
parameter [see eq.\ (\ref{min1})] and the masses of
the third family scalars.
The actual magnitude of the GUT effects depends on
the choice  of the free parameters
of the model. In section \ref{sec:s1} we introduced
a set of model-building parameters
\begin{equation}
m_{0},\,\, A_{0}, \,\, B_{0}, \,\, M_{1/2},
\label{basis1}
\end{equation}
where $B_{0}$ is traded for $\tan\beta$ using (\ref{min2}).
The parameters are  assumed  to be real
and $A_{0}$ (and $\mu$) can have either sign.
For fixed fermion masses [and SU(5) Yukawa couplings]
this set is enough, using renormalization group techniques,
to predict all the low-energy observables.
This is usually called up-down approach.
Applying a specific set  of values to these parameters
as a boundary condition at $M_{P}$ [$i.e.$, to eq.\ (\ref{rge2})]
or at $M_{G}$ [$i.e.$, to eq.\ (\ref{rge1})]
can lead to a quite different mass spectrum.
This is shown in the first and second
columns of Tables \ref{table:t1} -- \ref{table:t4}
for the scenarios illustrated
in Figs.\ \ref{fig:fig1} -- \ref{fig:fig4}.
LSP stands for the lightest supersymmetric particle which
is stable in the MSSM assuming $R$-parity.
Tables \ref{table:t1} -- \ref{table:t4}
give quantitative examples of the GUT corrections.
Note large deviations for observables that depend on
the $\mu$ parameter (such as
the Higgsino masses and components) and for the stop and sbottom masses.

Nevertheless, the free parameters of the model have to be extracted from
low-energy experiments. Thus, it is more convenient to choose a
set of free parameters of the model defined at $M_Z$, {\it i.e.},
bottom-up approach \cite{bottomup}.
For example, let us consider\footnote{Although the relation between
$m_{\tilde{e}_{L}}$ and $M_{\tilde{g}}$ and the
physical observables is straightforward, this is not the case for
$\tan\beta$ and $A_t(M_Z)$. Nevertheless, the latter two
can always be determined
as a function of physical observables such as the stop
or Higgs masses (these are usually  complicated functions involving
other parameters of the model).}
\begin{equation}
m_{\tilde{e}_{L}},\,\,  M_{\tilde{g}},\,\, \tan\beta,\,\, A_t(M_Z).
\label{basis2}
\end{equation}
This basis is also convenient because
the boundary conditions at $M_G$ can easily be obtained from
the parameters (\ref{basis2}).
Comparing now, for a given value
of (\ref{basis2}), the low-energy predictions
when the GUT effects are included (first column in
Tables \ref{table:t1} -- \ref{table:t4})
with  those when  the GUT effects are neglected (third column),
one finds more modest changes, especially in the $M_{1/2}>m_0$
region (Table \ref{table:t2}).
This is because using the basis (\ref{basis2}) one
eliminates global scalings of the SSB parameters arising in the
$M_P-M_G$ evolution (see Fig.\ \ref{fig:fig2} and comment above).
Although the choice (\ref{basis1}) is the relevant one
when speculating on the origin of the SSB parameters,
it will be$^{5}$ (\ref{basis2}) or a similar basis from which  the
value of the SSB parameters will be extracted once
supersymmetry is established.

Below, we study  in more detail the way in which
the predictions  from  naive $M_{G}$ universal boundary
conditions are smeared and modified by the non-universal GUT corrections.
Our aim is to study the uncertainties
and new regions of parameter space that could be opened
by those uncertainties. For example, we already mentioned
that a situation with $h_{b} > h_{t}$ can now be consistent
assuming pattern $(v)$.
We focus
on our choices $(a)$ and $(b)$ given in section \ref{sec:s1}.
Our numerical routines are similar to those described in
Ref.\ \cite{penn594}.
In short, we follow Ref.\ \cite{us556} in calculating the
couplings (and the unification point), and Ref.\ \cite{ep}
in treating the one-loop effective potential correction
$\Delta V$, including contributions from all sectors.
The boundary conditions
for $0 \leq m_{0} \leq 1000$ GeV,  $|A_{0}| \leq 3m_{0} $,
and $50 \leq M_{1/2} \leq 500$ GeV,
are picked at random, unless otherwise stated.
In order to minimize residual scale dependences (of order two-loop)
of the one-loop effective potential,
we rescale the Higgs potential (including wave function corrections)
to a typical $t$-scalar scale
of $600$ GeV before solving the one-loop minimization equations.
All Higgs masses include one-loop corrections calculated
using Ref.\ \cite{erz}; however, a ${\cal O}(10\%)$
ambiguity in the one-loop light Higgs boson mass remains \cite{penn594}.
We apply the conservative constraint $m_{h^{0}} \gtrsim 60 \pm 5$ GeV.
We also force all other relevant bounds on the mass parameters,
and require the correct EWSB ($i.e.$, a solution for $M_{Z}^{2}$),
a neutral LSP and positive squared masses for all physical scalars.
However, we do not minimize the full scalar potential in order to
eliminate color breaking minima that survive the
upper bound on $|A_{0}|$. That may affect the status of
some points with a particularly large
value of the $\mu$ parameter but is also sensitive to the choice
of $M_{G}$ or $M_{P}$.
For $\tan\beta = 42$ some points
could induce positive corrections to
the $b$-quark mass of more than $\sim 20\%$
and are omitted (smaller corrections could
be compensated by other threshold effects).
This effect is also
sensitive to the
$M_{G}$ or $M_{P}$ choice.


\subsection{First and second family scalars}
\label{sec:s31}

Since the Yukawa couplings of the first and second family of
squarks and sleptons are small, they can be neglected in the RGEs, {\it i.e.},
only the gauge contribution is relevant. The RGEs
can be solved analytically
and the physical scalar quark and lepton  ($\tilde f$) masses
in the basis~(\ref{basis1}) are given by
\begin{equation}
m^2_{\tilde f_{L,R}}=m^2_0+a_{\tilde f_{L,R}}M^2_{1/2}\pm M^2_Z\cos 2\beta[
T_{3f_{L,R}}-Q_{f_{L,R}}\sin^2\theta_W]+\Delta m^2_{\tilde f_{L,R}},
\label{fsg}
\end{equation}
where $a_{\tilde f_{L,R}}\sim 5-7$ for the squarks,  $\sim 0.5$ for the
left-handed sleptons
and $\sim 0.15$ for  the  right-handed sleptons.
$T_{3f_{L,R}}$ and $Q_{f_{L,R}}$, are the third component
of SU$(2)_L$ isospin and the electric charge of $f_{L,R}$, respectively.
The quantity
$\Delta m^2_{\tilde f_{L,R}}$  is the extra contribution  arising
 from the $M_{P}-M_G$ evolution and depends on the GUT.
For SU(5) we have \cite{us}
$\Delta m^2_{\tilde f_{L,R}}= [0.1a_{\tilde f_{L,R}} +
0.45]M^2_{1/2}$ for the left-handed
squarks, $\tilde u_R$
and $\tilde e_R$ and $\Delta m^2_{\tilde f_{L,R}}=
[0.1a_{\tilde f_{L,R}} + 0.3]M^2_{1/2}$
for the left-handed sleptons and $\tilde d_R$
(the term $0.1a_{\tilde f_{L,R}}$
arises due to the gaugino enhancement$^{4}$).
Note  that $\Delta m^2_{\tilde f_{L,R}}$
can be the dominant contribution to the $\tilde e_R$  mass and
can contribute $\sim 60\%$ to the $\tilde e_L$ and $\tilde\nu_L$
masses. For SO(10) D-terms, the magnitude of the GUT correction depends on
$M^2_D$. It is important to note that the requirement of non-tachionic
 sleptons, {\it i.e.}, $m_{\tilde e_{L,R}}>0$,
leads to strong constraints on $M^2_D$ from below and above that can be easily
obtained from eqs.~(\ref{so10}) and (\ref{fsg}).

When we use the input eq.\ (\ref{basis2}), however,  the scalar  masses read
\begin{equation}
m^2_{\tilde f_{L,R}}=m^2_{\tilde e_L}+a_{\tilde f_{L,R}}M^2_{\tilde g}
\pm M^2_Z\cos 2\beta[
T_{3f_{L,R}}-Q_{f_{L,R}}
\sin^2\theta_W]+0.3M^2_Z\cos 2\beta+\Delta m^2_{\tilde f_{L,R}},
\end{equation}
where now $a_{\tilde f_{L,R}}\sim 0.6-0.9$ for the squarks,  0 for the
left-handed sleptons
and $\sim -5\times 10^{-2}$ for  the  right-handed sleptons;
$\Delta m^2_{\tilde f_{L,R}}\sim
2\times 10^{-2}M^2_{\tilde g}$ for the left-handed squarks, $\tilde u_R$
and $\tilde e_R$ and $\Delta m^2_{\tilde f_{L,R}}=0$
for the left-handed leptons and $\tilde d_R$.
Thus, in the basis (\ref{basis2})
 GUT effects can change
 the $\tilde{e}_{R}$ (left-handed squarks and $\tilde{u}_{R}$)
masses by  $\sim 15\%$ $(1\%)$ at most.
An example is given in Table \ref{table:t2}.
The GUT effects in $m_{\tilde l}$
and  $m_{\tilde q}$ are
less apparent when  the basis (\ref{basis2})
is used instead of  the basis (\ref{basis1}).


\subsection{The $\mu$ parameter}
\label{sec:s32}

The $\mu$ parameter is extracted from the minimization condition
eq.\ (\ref{min1})
and depends on the values of $m^2_{H_i}$ at the weak scale.
If the quantities
$m^2_{H_i}$ are affected by the GUT effects, the $\mu$ parameter is
modified according to
\begin{equation}
\Delta\mu^2=-\frac{\Delta m^2_{H_1}-\Delta m^2_{H_2}\tan^2\beta}
{1-\tan^2\beta}\, ,
\label{shiftmu}
\end{equation}
where $\Delta m^2_{H_i}$ are the shifts in the soft Higgs masses
at the weak scale
due to the GUT effects.

\begin{figure}
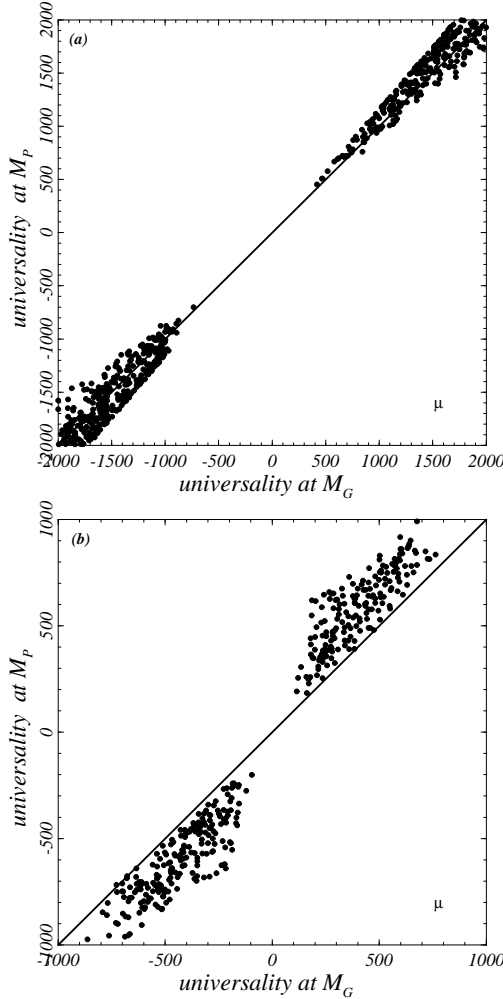

\caption{
The prediction for the Higgsino mass parameter $\mu$  (in GeV)
is compared assuming universality at $M_{G}$ and  at $M_{P}$ for
$(a)$ $m_t^{pole} = 160$ GeV, $\tan\beta = 1.25$ and for
$(b)$ $m_t^{pole} = 180$ GeV, $\tan\beta = 42$ (note the different scales).
$\lambda(M_{G}) = 1$, $\lambda^{'}(M_{G}) = 0.1$, and
the initial values for $m_{0}$, $A_{0}$ and  $M_{1/2}$ and the sign of $\mu$
are picked at random (see above).
}
\label{fig:fig5}
\end{figure}

\begin{figure}
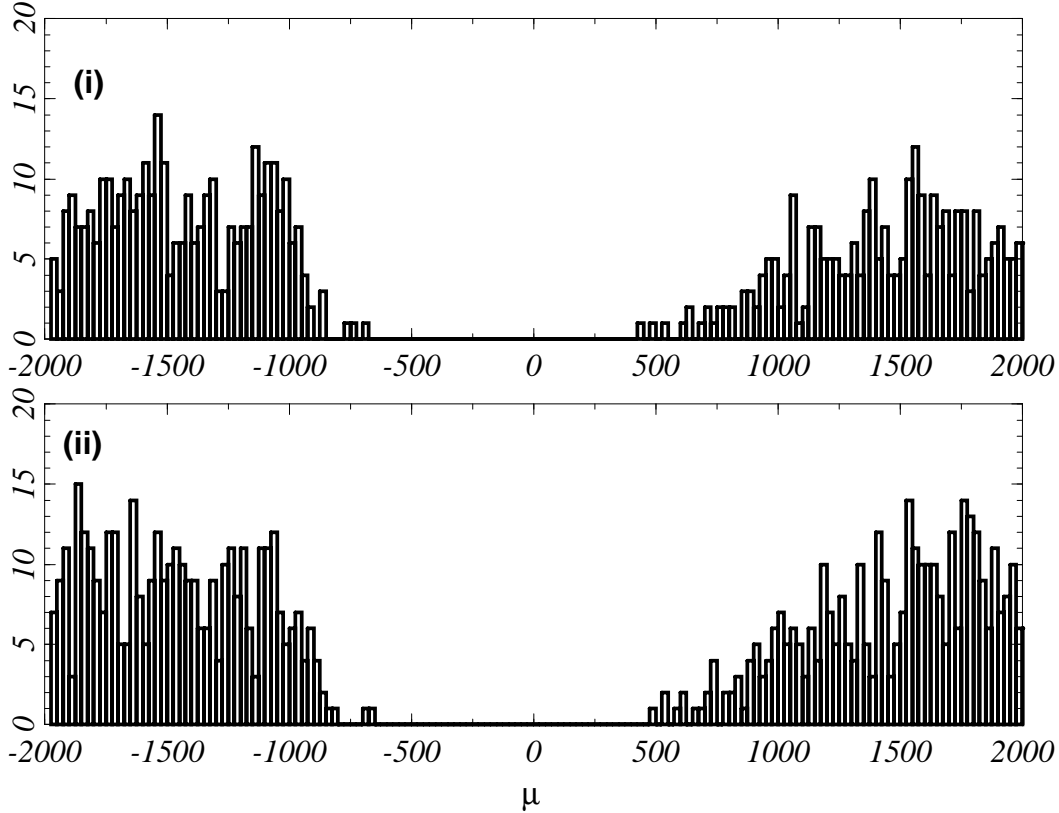

\caption{
The $\mu$ parameter prediction (in GeV)
in a sample of Monte Carlo calculations
for  $m_t^{pole} = 160$ GeV and $\tan\beta = 1.25$ and
assuming (i) universality at $M_{G}$ and (ii)
universality at $M_{P}$ and
$\lambda(M_{G}) = 1$, $\lambda^{'}(M_{G}) = 0.1$.
}
\label{fig:fig6}
\end{figure}

For low values of $\tan\beta$, one finds that $|\mu|$ is
typically large and not affected significantly by GUT corrections,
$\Delta\mu/\mu\sim 0.2$.
For the GUT pattern (${\it iii})$,
$\Delta\mu$ is small because $\Delta m^2_{H_2}$
is reduced at $M_Z$ (Fig.\ \ref{fig:fig3}).
For the  GUT pattern (${\it ii})$, this is
because of a partial cancellation between
$\Delta m^2_{H_1}$ and $\Delta m^2_{H_2}\tan^2\beta$.
In Fig.\ \ref{fig:fig5}a we compare the predictions of $\mu$
when universality is assumed at $M_P$ with
those when universality is assumed at $M_G$ for different
random points of the parameter
space defined by (\ref{basis1}).
The fact that $\Delta\mu$ depends on the sign of $\mu$
is due to the weak-scale threshold corrections to eq.\ (\ref{shiftmu})
that  can be substantial and have to be included.
In Fig.\ \ref{fig:fig6}
we show the distribution of the $\mu$ predictions
in a sample of Monte Carlo calculations. One can see that the distribution is
slightly changed. The differences in the
integrated area of the histograms give
an estimate of the changes in the allowed parameter space.

\begin{figure}
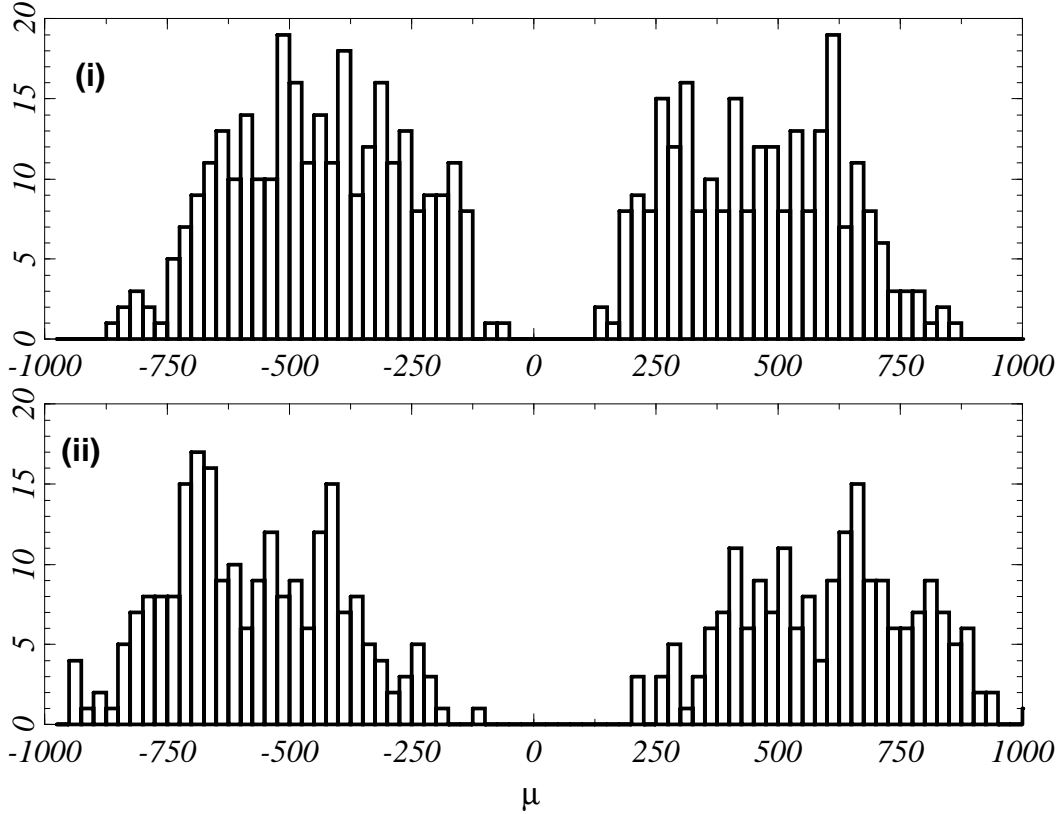

\caption{
Same as in Fig. 6 except
$m_t^{pole} = 180$ GeV and $\tan\beta = 42$.
}
\label{fig:fig7}
\end{figure}

For large values of $\tan\beta$, we have
$\Delta\mu^2\approx-\Delta m^2_{H_2}$.
For large $\lambda$, the splitting
$\Delta m^2_{H_2}$ can be substantial (if  $h_t$
is small, the quantity $\Delta m^2_{H_2}$ is
slightly  diminished in the $M_G-M_Z$ evolution -- see Fig.\ \ref{fig:fig4})
and $\mu$ can receive a large shift
(Figs.\ \ref{fig:fig5}b and \ref{fig:fig7}).
Note that the value of $|\mu|$ is
always increased since in the minimal SU(5) model $\Delta m^2_{H_2}<0$.

We have noted that
the minimal SU(5) effects lead generically to an
increase of $|\mu|$ [for low values of $\tan\beta$
the value of $|\mu|$ can be reduced (Fig.\ \ref{fig:fig5}a), but
$|\mu|$ is very large in
this regime and the effects are small].
When extended GUTs are considered, however,
this interesting feature is lost.
In order to have $\Delta\mu^2<0$ one needs $[\Delta m^2_{H_2}\tan^2\beta-
\Delta m^2_{H_1}]>0$,
and this latter condition can be obtained in the extended GUTs
considered in section \ref{sec:s2}.
In the large $\tan\beta$ regime, however,
one has also to consider the implications to EWSB.
{}From the minimization conditions of the
Higgs potential,
we have that EWSB requires at the weak scale (for large
$\tan\beta$),
\begin{equation}
\mu^2+m^2_{H_2}<0,\ \ \mu^2+m^2_{H_1}>0,
\label{equalities}
\end{equation}
which is difficult to achieve from universal $m^2_{H_i}=m^2_0$
at $M_G$ because $h_t\approx h_b$.
The GUT effects can produce a splitting
$\Delta m^2_{H_2}-\Delta m^2_{H_1}<0$ such that
eq.\ (\ref{equalities}) 
is satisfied
more easily. For example, in the GUTs considered
in section \ref{sec:s2} this splitting can be induced
but it requires $\Delta m^2_{H_2}<0$ which leads to
an increase of $|\mu|$.
In order to decrease $|\mu|$
we need $\Delta m^2_{H_2}>0$
that in the  extended GUTs considered can only
be obtained from the $D$--terms
eq.\ (\ref{so10}). In that case, however,
$\Delta m^2_{H_2}-\Delta m^2_{H_1}>0$
and  the EWSB is more difficult to obtain (requiring
even more fine-tuning). In fact, for a given point in
the parameter space there is an upper bound
on $\Delta m^2_{H_2}-\Delta m^2_{H_1}$
(that is strengthened when $\tan\beta$ increases).
Note that in cases in which (because of the $M_{P} - M_{G}$ evolution)
$m_{0}^{2} \gg m_{{\cal H}_{i}}^{2}$
the EWSB is even more difficult to obtain since $m_{H_{1}}^{2}$ and
$m_{H_{2}}^{2}$
can be both negative\footnote{A situation with both
$m_{H_{i}}^{2} +\mu^{2} < 0$ (when including loop corrections)
leads to an unacceptable minimum. Note that we plot
(in Figs.\ \ref{fig:fig1} -- \ref{fig:fig4})
the tree-level soft mass parameters.}
at $M_{Z}$.
Thus, GUT effects can  ease EWSB (for large $\tan\beta$) but,
in that case, they increase the value of $|\mu|$.

The fact that $\mu$ is extracted from eq.\ (\ref{min1}) and is
usually larger than $M_Z$ implies that the
lightest chargino ($\chi^+_1$) and neutralinos ($\chi^0_1$ --
the LSP -- and $\chi^{0}_{2}$)
are mostly gauginos. As we have shown, this
property is not altered by GUT effects of the minimal SU(5) model.
Therefore, the masses of $\chi^+_1$ and
$\chi^0_{1,2}$ depend mostly on $M_{1/2}$ and are almost independent
of the soft scalar masses. The $M_P-M_G$ evolution can
enhance $M_5$ and thus the gaugino masses by $\sim 10\%$.
Using the basis (\ref{basis2}), however, the gaugino masses can be
written as a function of  $M_{\tilde g}$, {\it i.e.}, independent
of the scale.


\subsection{The Higgs scalars}
\label{sec:s33}

The masses and mixing angles of the
Higgs bosons in the MSSM can be written as a function
of two parameters that we choose to be $\tan\beta$ and the
pseudoscalar Higgs mass $m^2_{A^0}\equiv 2\mu^2+m^2_{H_1}+m^2_{H_2}$.
Since $\tan\beta$ is considered an input parameter, only
the GUT effects in $m^2_{A^0}$ are relevant.
A shift in the soft Higgs masses arising
from GUT physics  shifts
$m^2_{A^0}$ by
\begin{equation}
\Delta m^2_{A^0}=\frac{1+\tan^2\beta}{1-\tan^2\beta}
(\Delta m^2_{H_2}-\Delta m^2_{H_1})\, .
\end{equation}
For $\tan\beta\approx 1$, the behavior of $\Delta m^2_{A^0}$ is similar
to  that of $\Delta\mu^2$ discussed in the previous section.
For large $\tan\beta$, one
has $\Delta m^2_{A^0}\approx\Delta m^2_{H_1}-\Delta m^2_{H_2}$.
Since in the minimal SU(5) model  $\Delta m^2_{H_2}
\approx \Delta m^2_{H_1}$, one has $\Delta m^2_{A^0}\approx 0$.

The mass of the lightest Higgs $h^0$  receives large radiative correction
induced by  loops involving the top and stop ($m_{h^{0}} \rightarrow
m_{h^{0}} + \Delta_{h^{0}}[m_{t},m_{Q},m_{U},A_{t},\mu,\tan\beta]$)
and can be
changed if either the diagonal or off-diagonal entries
in the stop mass matrix are shifted by GUT effects
(section \ref{sec:s34}).
The effects are negligible for $\tan\beta \gtrsim 2$ where
the Higgs boson is heavy at tree level.
However, for a light tree-level Higgs boson ($\tan\beta \approx 1$),
unless the different effects cancel (see Tables \ref{table:t1}
and  \ref{table:t2}), they can modify
$m_{h^{0}}$ by a few GeV (see Table \ref{table:t3}).
The cancellations depend on the sign of the $\mu$ parameter.
In Figs.\ \ref{fig:fig8} and \ref{fig:fig9} we show the distribution
of the lightest Higgs mass.
One can note that the distributions are only slightly sensitive
to the GUT corrections so that
previous calculations ($e.g.$, see \cite{penn594})
of the predictions of $m_{h^0}$ in SUSY GUTs are not altered.
(Fig.\ \ref{fig:fig8}, here, roughly corresponds to Figs.\ 9a and 10a
in Ref. \cite{penn594}.)

When GUT effects from extended models are considered, the value of $m^2_{A^0}$
can increase (decrease) if a splitting
$\Delta m^2_{H_1}-\Delta m^2_{H_2}>0$ $(<0)$ is induced.
Again, when  EWSB and non-tachionic particles are required,
$\Delta m^2_{A^0}$ can be bounded from below and above.


\subsection{Third family scalars}
\label{sec:s34}

As  explained above,
the masses of the scalars of the third family can be largely modified
by the GUT effects due to  a large $h_t$\footnote{There are also
gauge GUT effects that
are the same as those to the first and second family scalars.}.
The  GUT effects can modify
(i)  the soft mass $m_{10}^{2}$,
(ii)  $m_{H_2}^{2}$
and hence the evolution of $m_{Q}^{2}$ and  $m_{U}^{2}$,
and (iii)  the $\mu$ parameter (section \ref{sec:s32})
and  $A_{t,b,\tau}$ that enter in the left-right mixing term of the scalar
masses. These three effects compete with each other to increase or decrease
the masses.

The lightest stop, $\tilde t_1$, has recently received much attention.
Its mass, $m_{\tilde t_1}$,
 is usually smaller than the mass of the other squarks and
can induce significant one-loop effects in
low-energy processes
such as $Z\rightarrow b\bar b$ and $b\rightarrow s\gamma$.
In the minimal SU(5) model, the dominant GUT effect in $m_{\tilde t_1}$
arises from (ii). Since
$m^2_{H_2}$ is diminished by GUT effects, $m^{2}_{Q}(M_{Z})$
and $m^{2}_{U}(M_{Z})$ are larger.
Thus, we find that $m_{\tilde t_1}$ is always enhanced.
It follows that
some points of the parameter space which correspond to a tachionic
$t$-scalar and are excluded  when  the $M_{P}$ to $M_{G}$ evolution is
neglected, can be allowed.
In Fig.\ \ref{fig:fig10}
we compare the predictions
of  $m_{\tilde t_1}$ with and without the $M_P-M_G$ evolution.
We find significant corrections in the low $\tan\beta$ regime, especially
for small values of  $m_{\tilde t_1}$.
This implies that
one-loop corrections induced by  $\tilde t_1$ to low-energy processes
can be significantly reduced.
Note that $m_{\tilde{t}_{1}} \gtrsim 200$ GeV when including
the $M_{P} - M_{G}$ evolution.

The GUT effects from non-minimal SU(5) models [pattern ({\it v})] usually
 enhance $m_{\tilde t_1}$
since (ii) is still dominant. In the minimal SO(10), however,
the splittings eq.\ (\ref{so10}) can lead to a lighter stop.
For  $M^2_D<0$, one has $\Delta m^2_{10}<0$  and $\Delta m^2_{H_2}>0$, and
both effects (i) and (ii) decrease  $m_{\tilde t_1}$.
 Since, as we said in
section \ref{sec:s31},
these splittings can lead to tachionic sleptons,  $m_{\tilde t_1}$
cannot  be reduced significantly.

For  the sbottom and stau we  find that the effects (i) and (ii)
can be equally important and their masses
can increase or decrease depending on the point in the parameter space
(see tables).


\subsection{Possible implications}
\label{sec:s35}

To conclude our survey  of the weak-scale phenomena,
we summarize the most interesting implications of the GUT effects
for experiment.
We have shown that assuming the minimal SU(5) model and
universality at $M_{P}$ instead of $M_{G}$ one typically predicts heavier
particles.
For example, the scalar
leptons could be substantially heavier (see Table \ref{table:t2}).
More interestingly, correlations between the different
parameters are modified (see also \cite{us}), $i.e.$, correlations calculated
assuming universality at $M_{G}$ could be misleading and should
not be used to constrain the parameter space.
Smeared correlations imply less EWSB-related fine-tuning
and that a larger parameter space may be available.
For example, the $\tilde{t}_{1}$ mass is shown
[for choice $(a)$] in
Figs.\ \ref{fig:fig11} and \ref{fig:fig12}.
It is typically larger when considering
the GUT effects and
its correlation with the $\chi_{1}^{+}$ mass
(or for that matter, with the gluino mass -- see Fig.\ 3 of Ref. \cite{us})
is smeared while that with the
$\tilde{t}_{2}$ mass
is strengthened. For choice $(b)$ the $\tilde{t}$ masses
are only slightly altered but
$\chi_{1}^{+}$ could be heavier. In Fig.\ \ref{fig:fig13}
we examine the Higgsino
fraction of the LSP for choice $(b)$, which is relevant, $e.g.$,
for relic abundance calculations. The larger Higgsino mass implies
smaller Higgsino fractions of the gaugino-like
$\chi^{+}_{1}$,  $\chi^{0}_{1}$ and  $\chi^{0}_{2}$.
That and the heavier $\tilde{t}_{1}$ lead to a stronger decoupling
of the supersymmetric particles from
low-energy processes, $e.g.$, from $Z\rightarrow b\bar{b}$.

In extended GUT models the restrictions on the parameter space
from EWSB are somewhat relaxed. In particular, values of
$m_{t}^{pole}$ and $\tan\beta$ which imply $h_{b} > h_{t}$
may be consistent with EWSB. Non-vanishing $D$-terms can lead to
a very different spectrum
in comparison
to the situation with vanishing $D$-terms.
For example, they could
lead to a lighter Higgsino which, as discussed above, is an
interesting possibility phenomenologically. However,
when combined with
the $M_{P} - M_{G}$ evolution, the effects are diminished
(see Table \ref{table:t4}).
Also, in the models studied
the amount that the $\mu$ prediction can be diminished by GUT effects
is strongly constrained
(unlike in some ad hoc cases studied in Ref. \cite{o1,o2,o3,o4}).
We conclude that
if able to observe the 
GUT effects,
$e.g.$, from correlation measurements,
collider experiments could directly probe the GUT scale physics.
That is a non-trivial task and it would depend
on the experimental resolution as well as on the
region of parameter space nature chooses.


\section{Conclusions}
\label{sec:s4}

Above, we examined the effects of a grand-unified symmetry
between the Planck and GUT scales in the SSB parameters.
Our only assumptions were coupling constant unification
at $M_{G} \approx 2 \times 10^{16}$ GeV; the MSSM as the effective
theory below that scale; and universal SSB parameters at
the minimal supergravity scale $M_{P} \approx 2 \times 10^{18}$ GeV.
In addition, we had to specify the GUT.
We previously analyzed these assumptions in Ref. \cite{us}
assuming the minimal SU(5) model and found potentially
large deviations from universality for the SSB parameters at $M_{G}$.
In particular, we emphasized the role of large Yukawa couplings
which are generic in such models.
Here, we further cataloged the possible patterns of non-universality
in that model and examined in great detail their implications to the weak
scale phenomena.
We found potentially large corrections (in comparison with the working
assumption of universality at $M_{G}$)
to the allowed
parameter space, the $\mu$ parameter and to the third-family
squark spectrum.
These are all related primarily to
the $M_{P} - M_{G}$ evolution of the light Higgs fields.
In the gaugino-dominated cases ($e.g.$, in no-scale models)
large corrections to the scalar-lepton masses are also possible.
The Higgs and the first and second family squark
sectors are relatively insensitive to the GUT effects.
A different situation may arise in extended models. For example,
we discussed the situation in
non-minimal SU(5), where $m_{H_{1}}^{2}(M_{G})$ and $m_{H_{2}}^{2}(M_{G})$
are independent parameters,
as well as  the appearance of non-vanishing
$D$-terms in SO(10).
In both models correlations are further diminished and
EWSB constraints are  more easily satisfied.
However, EWSB still plays an important role in constraining
GUT effects, $e.g.$, effects that could diminish $\mu$.
Implications of the above
to experiment were already summarized in section \ref{sec:s35}.

Finally, let us comment on
the predictive power of the MSSM. Assuming minimal SU(5),
two additional parameters are needed (only one of which
plays an important role). In non-minimal SU(5),
SU(5)$\times$U(1) and SO(10) three or more new
parameters are needed. The more parameters
the model has, the larger role GUT effects
could play in weak-scale phenomena, but the less predictive is the model.
On the other hand, when adding only a small number of new parameters,
the effects in different SSB parameters are correlated, and thus,
constrained ($e.g.$, by EWSB).
The predictive power can be  further altered when considering
threshold corrections. These are described in detail
for the minimal SU(5) model in Appendix \ref{sec:a2}
[eqs.\ (\ref{oneloopmata})-- (\ref{lasteq})]
and, in general, they do not significantly modify the
tree-level patterns described in section \ref{sec:s2}.
In extended models threshold corrections could be more important
if more and larger representations are present. Also, perturbation
theory could break down in these models and one would need
non-perturbative methods to calculate
the GUT effects.

GUT effects in the SSB parameters are generic, leading to non-universal
patterns different than those, $e.g.$, in string theory,  and
could probe the GUT-scale physics. However, they are model-dependent
and lead to uncertainties in any model-independent
analysis, which typically assumes universality at $M_{G}$.
Until supersymmetry is
established and characterized,
the effects have
to be considered as uncertainties to supergravity GUT model predictions.

\acknowledgments
It is a pleasure to thank Paul Langacker for discussions and comments
on the manuscript.
This work was supported by the US Department of Energy
Grant No. DE-AC02-76-ERO-3071 (NP) and by the Texas Commission
Grant No. RGFY93-292B (AP).

\appendix


\section{The minimal SU(5) model}
\label{sec:a1}

The Higgs sector of the model consists of three supermultiplets,
$\Sigma({\bf 24})$ in the adjoint representation [which is
responsible for the breaking of SU(5) down to
SU(3)$_c\times$SU(2)$_L\times$U(1)$_Y$],
$ {\cal H}_{1} ({\bf\bar 5})$ and ${\cal H}_{2} ({\bf 5})$:
\begin{equation}
\Sigma\equiv\sqrt{2}T_{a}w_{a}\, \ ,\ \ \
{\cal H}_{1}=\left(\matrix{H_{C_1}\cr
H_1}\right)\, \ ,\ \ \ \ {\cal H}_{2}=\left(\matrix{H_{C_2}\cr
H_2}\right)\, ,
\end{equation}
where $H_{C_i}$ and $H_i$ are the color triplets and SU(2)$_L$ doublets,
respectively,  and $T_a$  are the SU(5) generators with
 ${\rm tr}\{T_aT_b\}=\delta_{ab}/2$.
The matter superfields are in the ${\bf\bar 5}+{\bf 10}$ representations,
$\phi({\bf\bar 5})$ and $\psi({\bf 10})$.
The superpotential is given by
\begin{eqnarray}
W&=&\mu_\Sigma{\rm tr}\Sigma^{2}+\frac{1}{6}\lambda^{'}{\rm tr}\Sigma^{3}
+\mu_H {\cal H}_{1}{\cal H}_{2} +\lambda
{\cal H}_{1}\Sigma {\cal H}_{2}\nonumber \\
&+&\frac{1}{4}h_t\epsilon_{ijklm}\psi^{ij}\psi^{kl}{\cal H}_{2}^m
+\sqrt{2}h_b\psi^{ij}\phi_{i}{\cal H}_{1j}\, ,
\label{super}
\end{eqnarray}
where we have omitted family indices and $h_t$ and $h_b$
are the Yukawa couplings of the third generation (we neglect
the other Yukawa couplings).
In the supersymmetric limit
$\Sigma$ develops a
vacuum expectation value
$\langle\Sigma\rangle =\nu_{\Sigma}\, {\rm diag}
(2,2,2,-3,-3)$ and
the  gauge bosons  $X$ and $Y$ receive a mass
$M_V=5g_{G}\nu_{\Sigma}$.
In order for
the Higgs SU(2) doublets to have masses  of ${\cal O}(M_Z)$ instead
of ${\cal O}(M_G)$,
the fine-tuning
$\mu_H-3\lambda \nu_{\Sigma} \lesssim {\cal O}(M_Z)$
is required and one obtains
$M_{H_{C}}=\frac{\lambda}{g_G} M_V$.
Dimension-five operators
induced by the color triplets give large contributions
$\propto 1/M^2_{H_C}$
to the proton decay rate \cite{proton}.
To suppress such operators, the mass of the color triplets has to
be large, $M_{H_{C}}\gtrsim M_V$,
implying
$\lambda\gtrsim g_G\approx 0.7$.

Below $M_P$, the effective lagrangian also contains the SSB terms (note
that we keep the same notation for the superfields and their corresponding
scalar fields)
\begin{eqnarray}
-{\cal L}_{soft}&=&m^2_{ {\cal H}_{1}}| {\cal H}_{1}|^2+
m^2_{{\cal H}_{2}}|{\cal H}_{2}|^2
+m^2_{\Sigma}{\rm tr}\{\Sigma^{\dag}\Sigma\}
+m^2_5|\phi|^2+m^2_{10}{\rm tr}\{\psi^{\dag}\psi\} \nonumber \\
&+&[B_\Sigma \mu_\Sigma{\rm tr}\Sigma^2
+\frac{1}{6}A_{\lambda^{'}}
\lambda^{'}{\rm tr}\Sigma^3
+B_H\mu_H {\cal H}_{1}{\cal H}_{2}
+A_{\lambda}\lambda {\cal H}_{1}\Sigma{\cal H}_{2}\nonumber \\
&+&\frac{1}{4}A_th_t\epsilon_{ijklm}\psi^{ij}\psi^{kl}{\cal H}_{2}^m
+\sqrt{2}A_bh_b\psi^{ij}\phi_{i}{\cal H}_{1j}+
\frac{1}{2}M_5\lambda_{\alpha}
\lambda_{\alpha}+h.c.],
\label{soft}
\end{eqnarray}
where $\lambda_{\alpha}$ are the gaugino fields.

The SU(5) RGEs for the
SSB parameters and Yukawa couplings
are given by
\begin{eqnarray}
\frac{dm^2_{10}}{dt}&=&\frac{1}{8\pi^2}
[3h^2_t(m^2_{{\cal H}_{2}}+2m^2_{10}
+A^2_t)+2h^2_b(m^2_ {{\cal H}_{1}}+m^2_{10}+m^2_{5}+A^2_b)
-\frac{72}{5}g_G^2M_{5}^2]\, ,\nonumber \\
\frac{dm^2_{5}}{dt}&=&\frac{1}{8\pi^2}
[4h^2_b(m^2_{ {\cal H}_{1}}+m^2_{10}
+m^2_{5}+A^2_b)-\frac{48}{5}g_G^2M_{5}^2]\, ,\nonumber \\
\frac{dm^2_{ {\cal H}_{1}}}{dt}&=&\frac{1}{8\pi^2}
[4h^2_b(m^2_{ {\cal H}_{1}}+m^2_{10}
+m^2_{5}+A^2_b)
+\frac{24}{5}\lambda^2(m^2_{ {\cal H}_{1}}+m^2_{{\cal H}_{2}}+
m^2_{\Sigma}+A^2_{\lambda})
-\frac{48}{5}g^2_GM_{5}^2]\, ,\nonumber \\
\frac{dm^2_{{\cal H}_{2}}}{dt}&=&\frac{1}{8\pi^2}
[3h^2_t(m^2_{{\cal H}_{2}}+2m^2_{10}
+A^2_t)
+\frac{24}{5}\lambda^2(m^2_{ {\cal H}_{1}}+
m^2_{{\cal H}_{2}}+m^2_{\Sigma}+A^2_{\lambda})
-\frac{48}{5}g^2_GM_{5}^2]\, ,\nonumber \\
\frac{dm^2_{\Sigma}}{dt}&=&\frac{1}{8\pi^2}[\frac{21}{20}
\lambda^{\prime 2}
(3m^2_{\Sigma}+A^2_{\lambda '})
+\lambda^2(m^2_{ {\cal H}_{1}}+m^2_{{\cal H}_{2}}
+m^2_{\Sigma}+A^2_\lambda)
-20g^2_GM_{5}^2]\, ,\nonumber \\
\frac{dA_t}{dt}&=&\frac{1}{8\pi ^2}[9A_th_t^2+4A_bh_b^2
+\frac{24}{5}A_\lambda \lambda ^2
-\frac{96}{5}g^2_GM_{5}]\, ,\nonumber \\
\frac{dA_b}{dt}&=&\frac{1}{8\pi ^2}[10A_bh_b^2+3A_th_t^2
+\frac{24}{5}A_\lambda \lambda ^2
-\frac{84}{5}g^2_GM_5]\, ,\nonumber\\
\frac{dA_\lambda }{dt}&=&\frac{1}{8\pi ^2}[\frac{21}{20}A_{\lambda '}
\lambda^{\prime 2}+
3A_th_t^2+4A_bh_b^2
+\frac{53}{5}A_\lambda
\lambda ^2-\frac{98}{5}g^2_GM_5]\, ,\nonumber \\
\frac{dA_{\lambda '}}{dt}&=&\frac{1}{8\pi ^2}[\frac{63}{20}
A_{\lambda '}\lambda^{\prime 2}+3A_\lambda\lambda^2-30g^2_GM_5]\, ,
\nonumber \\
\frac{dh_t}{dt}&=&\frac{h_t}{16\pi ^2}[9h^2_t+4h^2_b
+\frac{24}{5}\lambda^2-\frac{96}{5}g^2_G]\, ,\nonumber \\
\frac{dh_b}{dt}&=&\frac{h_b}{16\pi ^2}[10h^2_b+3h^2_t+\frac{24}{5}
\lambda^2
-\frac{84}{5}g^2_G]\, ,\nonumber \\
\frac{d\lambda }{dt}&=&\frac{\lambda }{16\pi ^2}[\frac{21}{20}
\lambda^{\prime 2}+3h^2_t+4h^2_b+\frac{53}{5}\lambda ^2
-\frac{98}{5}g^2_G]\, ,\nonumber \\
\frac{d\lambda '}{dt}&=&\frac{\lambda '}{16\pi ^2}[\frac{63}{20}
\lambda^{\prime 2}+3\lambda^2-30g^2_G]\, ,
\label{rgesu5}
\end{eqnarray}
where $t=\ln {\cal Q}$.
The RGE for the gauge coupling is
$d\alpha_G/dt=-3\alpha_G^2/2\pi$, and
similarly $dM_{5}/dt=-3\alpha_G M_{5}/2\pi$.
We can omit the RGEs for $\mu_\Sigma$, $\mu_H$, $B_\Sigma$ and $B_H$,
which are arbitrary parameters that decouple from the rest
of the RGEs.

Below $M_G$, the effective theory  corresponds to the MSSM:
\begin{equation}
W=\mu H_1H_2+h_tQH_2U+h_bQH_1D+h_{\tau}LH_1E,
\label{effsuper}
\end{equation}
where $Q$ and $L$ are, respectively,
the quark and lepton SU(2)$_L$ doublets, and
$U$, $D$ and $E$ are, respectively, the quark and lepton
SU(2)$_L$ singlets.


\section{GUT-scale threshold corrections}
\label{sec:a2}

Even if the scale where universal SSB terms are generated is assumed to be
$M_G$, there is some arbitrariness in the value of $M_G$ due to the
mass-splitting
between  the particles at the GUT scale, {\it i.e.}, threshold effects.
These GUT effects to the SSB terms (to the best of our knowledge)
have never  been considered before.
As we will show,  they can be as important as the low-energy
(supersymmetric) threshold effects,
which are the only threshold effects to the SSB parameters
considered in the literature.

We will only consider GUT threshold corrections to the scalar
SSB parameters. Corrections to the gaugino masses have been
computed in Ref.\ \cite{gaugino}, where they were shown to be small.
There are two ways to compute the threshold correction to the scalar SSB
parameters. One way consists of
 calculating explicitly the one-loop
diagrams that contribute
to the scalar SSB terms.
A second way, which is much
 simpler, consists of obtaining the one-loop SSB terms from the
 one-loop effective potential that in the Landau gauge and
in the dimensional-reduction ($\overline{\mbox{DR}}$) scheme reads
\begin{equation}
\Delta V=\frac{1}{64\pi^2}\sum_i(-1)^{2s_i}(2s_i+1)
 M_i^4\left[\ln\frac{M_i^2}{{\cal Q}^2}-
\frac{3}{2}\right]\, ,
\label{effpo}
\end{equation}
where $M_i^2$ and $s_i$ are, respectively,  the field-dependent
squared mass and  spin of the particle $i$. In this case,
one only  has to compute the masses $M_i$.
We will present the GUT threshold correction using (\ref{effpo})
although we have  checked the results with
 those from the explicit diagramatic
calculation.

Let us start with the one-loop correction to the
SSB squared-mass  $m^2_{i}$ of a scalar $\Phi_i$ induced by
a heavy chiral supermultiplet, which consists of
a fermion field with mass $M_F$ and
two (real) scalars with masses $M_{S_1}$ and  $M_{S_2}$. In the
supersymmetric limit $M_F=M_{S_{1,2}}\equiv M_0$ where $M_0$,
is of the order $M_G$.
We separate the GUT effects into logarithmic corrections and finite
corrections:

\noindent 1. Logarithmic corrections: The
 logarithmic term  of eq.~(\ref{effpo}) gives
 the one-loop  contribution to $m^2_{i}$
\begin{eqnarray}
\Delta m^2_{i}(log)&=&\frac{1}{32\pi^2}\left\{\left[
\frac{\partial M^2_{S_1}}{\partial |\Phi_i|^2}M^2_{S_1}
+\frac{\partial M^2_{S_2}}{\partial |\Phi_i|^2}M^2_{S_2}
-2\frac{\partial M^2_{F}}{\partial |\Phi_i|^2}M^2_{F}\right]
\ln\frac{M^2_F}{{\cal Q}^2}\right.\label{loga}\\
&+&\left.\frac{\partial M^2_{S_1}}{\partial |\Phi_i|^2}
M^2_{S_1}\ln\frac{M^2_{S_1}}{M^2_F}
+\frac{\partial M^2_{S_2}}{\partial |\Phi_i|^2}M^2_{S_2}
\ln\frac{M^2_{S_2}}{M^2_F}\right\}\,  .
\label{logb}
\end{eqnarray}
The first term (\ref{loga}) gives the logarithmic contribution
arising from the energy-scale difference between the  mass of
the superfield and the scale
${\cal Q}$ (splittings between  different heavy superfields).
The second term (\ref{logb}) arises from  the boson-fermion mass
splitting within
the superfield.
 This latter type of correction  to the
 Yukawa and gauge couplings
 is of ${\cal O} (m^2_{soft}/M^2_0)$ and then
negligible for $M_0\approx M_G\gg m_{soft}$. However, it can be important
to the $m^2_i$, as we will show below.

\noindent 2. Finite corrections: The non-logarithmic contribution to
$m^2_i$ from (\ref{effpo}) is given by
\begin{equation}
\Delta m^2_{i}(finite)=\frac{-1}{32\pi^2}\left[\frac{\partial M^2_{S_1}}
{\partial |\Phi_i|^2}M^2_{S_1}
+\frac{\partial M^2_{S_2}}{\partial |\Phi_i|^2}M^2_{S_2}
-2\frac{\partial M^2_{F}}{\partial |\Phi_i|^2}M^2_{F}\right]\, .
\label{finite}
\end{equation}
Note that this contribution depends on the renormalization scheme.
Eq.~(\ref{finite}) has been obtained in the $\overline{\mbox{DR}}$ scheme.
The heavy squared-masses $M^2_{S_{1,2}}$ and $M^2_F$ can be
written using a $1/M_0$ expansion as
\begin{eqnarray}
M^2_{S_{1,2}}&=&M^2_0\pm aM_0+b+\sum_i\left[
b_i\pm\frac{c_i}{M_0}+\frac{d_i}{M^2_0}\right]|\Phi_i|^2+\dots\, ,\nonumber\\
M^2_F&=&M^2_0+\sum_i b_i|\Phi_i|^2+\dots\, ,
\label{massl}
\end{eqnarray}
where the coefficients $a$--$d_i$ depend on the SSB parameters\footnote{
The coefficients $a$--$d_i$ also depend on the mass parameters of the
superpotential (such as the $\mu$ parameter), but this dependence
has to be discarded since we are only interested in the
corrections to
the SSB parameters.},
and we have only kept the
relevant terms for our analysis.
Substituting  eq.~(\ref{massl}) in eqs.~(\ref{loga}), (\ref{logb})
and (\ref{finite}),
 we get
\begin{eqnarray}
\Delta m^2_{i}(log)&=&
\frac{1}{16\pi^2}[bb_i+ac_i+d_i]\ln\frac{M^2_F}{{\cal Q}^2}\label{corra}\\
&+&\frac{1}{32\pi^2}[b_i(2b+a^2)+2ac_i]+\dots\, ,\label{corrb}\\
\Delta m^2_{i}(finite)&=&\frac{-1}{16\pi^2}[bb_i+ac_i+d_i]+\dots
\, .\label{corrc}
\end{eqnarray}
Notice that the contribution from (\ref{logb}) [the term (\ref{corrb})]
 gives a correction to $m^2_i$
 not suppressed by powers of  ${\cal O} (m^2_{soft}/M^2_0)$,
although it turns out to be  non-logarithmic.
{}From eqs.~(\ref{corra})--(\ref{corrc}), the GUT-scale corrections can
easily be obtained if
the dependence of the heavy masses on the light scalar fields
 (the $a$--$d_i$ coefficients) is known.
In the minimal SUSY SU(5) model (see Appendix A),
 the GUT spectrum consists of the 12 vector
superfields $V=X,Y$, the color triplets $H_{C_i}$, and the
$\Sigma$ superfield. With respect to
SU(3)$_c\times$SU(2)$_L$,
the $\Sigma$ supermultiplet decomposes
 into $(3,2)+(\bar 3,2)+(8,1)+(1,3)+(1,1)$. In the supersymmetric limit,
the $(3,2)$ and $(\bar 3,2)$ components are degenerate with $X$ and $Y$.
The $(1,3)$ and $(8,1)$ components, $\Sigma_{3}$ and $\Sigma_{8}$,
respectively,
 have a common  mass $10\mu_{\Sigma}$, while the mass
of the singlet, $\Sigma_{1}$, is $2\mu_{\Sigma}$.
 When the SSB terms are considered,
a boson-fermion mass splitting within every supermultiplet is induced.

Considering only
 the two large Yukawa couplings,
$\lambda$ and $h_t$, the coefficients  $a$--$d_i$
for the  $\Sigma_3$ are given by
\begin{eqnarray}
a&=&B_{\Sigma}\, ,\ \ \ b=m^2_{\Sigma}\, ,\nonumber\\
b_{H_1}&=&b_{H_2}=\lambda^2\, ,\nonumber\\
c_{H_1}&=&c_{H_2}=\frac{\lambda^2}{2}[2A_{\lambda}-B_{\Sigma}]\, ,\nonumber\\
d_{H_1}&=&\frac{\lambda^2}{2}[m^2_{{\cal H}_{2}}-m^2_{\Sigma}+A_{\lambda}^2
+B_{\Sigma}^2
-2A_{\lambda}B_{\Sigma}]\, ,\nonumber\\
d_{H_2}&=&\frac{\lambda^2}{2}[m^2_{{\cal H}_{1}}-m^2_{\Sigma}+A_{\lambda}^2
+B_{\Sigma}^2
-2A_{\lambda}B_{\Sigma}]\, .
\label{sigmat}
\end{eqnarray}
The coefficients for $\Sigma_1$ can be obtained from eqs.~(\ref{sigmat})
by replacing
 $\lambda^2\rightarrow \frac{3}{5}\lambda^2$.
For the color triplets $H^{\alpha}_{C_i}$ ($\alpha$ being the color index)
we have,   assuming $M_{H_{C}}> M_V$,
\begin{eqnarray}
a&=&B_{H}\, ,\ \ \ b=(m^2_{{\cal H}_{1}}+m^2_{{\cal H}_{2}})/2\, ,\nonumber\\
b_{H_1}&=&b_{H_2}=\lambda^2\, ,\nonumber\\
c_{H_1}&=&c_{H_2}=\frac{\lambda^2}{2}[2A_{\lambda}-B_{H}]\, ,\nonumber\\
d_{H_1}&=&\frac{\lambda^2}{2}[m^2_{\Sigma}-m^2_{{\cal H}_{1}}+A_{\lambda}^2
+B_{H}^2
-2A_{\lambda}B_{H}]\, ,\nonumber\\
d_{H_2}&=&\frac{\lambda^2}{2}[m^2_{\Sigma}-m^2_{{\cal H}_{2}}+A_{\lambda}^2
+B_{H}^2
-2A_{\lambda}B_{H}]\, ,\nonumber\\
b_{U_{\alpha}}&=&b_{Q_{\beta}}=b_{E}=h_t^2\, ,\ \ \ \ \ \beta\not=
\alpha\, ,\nonumber\\
c_{U_{\alpha}}&=&c_{Q_{\beta}}=c_{E}=\frac{h_t^2}{2}
[2A_{t}-B_{H}]\, ,\nonumber\\
d_{U_{\alpha}}&=&d_{Q_{\beta}}=d_{E}=\frac{h_t^2}{2}
[m^2_{10}-m^2_{{\cal H}_{1}}+A_{t}^2
+B_{H}^2-2A_{t}B_{H}]\, .
\label{colort}
\end{eqnarray}
The rest of the heavy fields decouple from the light scalar fields
(gauge contributions have not
 been considered).

There are also one-loop corrections coming from the wave-function
renormalization constants that cannot be obtained from
the one-loop effective potential.  These corrections have to
be calculated from the explicit one-loop diagrams, and give a contribution
to the Higgs soft masses
\begin{equation}
m^2_{H_{i}}({\cal Q})=m^2_{{\cal H}_i}({\cal Q})+
\frac{\lambda^2}{8\pi^2}m^2_{{\cal H}_i}
\left[\frac{3}{4}\ln\frac{M^2_{\Sigma_{3}}}{{\cal Q}^2}
+\frac{3}{20}\ln\frac{M^2_{\Sigma_{1}}}{{\cal Q}^2}
+\frac{3}{2}\ln\frac{M^2_{H_C}}{{\cal Q}^2}-\frac{12}{5}\right]\, ,
\label{wfrca}
\end{equation}
and to the sleptons and squarks soft masses
\begin{equation}
 m^2_i({\cal Q})=m^2_{10}({\cal Q})+\frac{n_ih_t^2}{16\pi^2}m^2_{10}
\left[\ln\frac{M^2_{H_C}}{{\cal Q}^2}-1\right]\, ,
\label{wfrcb}
\end{equation}
where $n_i=1,2,3$ for $i=U,Q,E$, respectively.
Inserting eqs.~(\ref{sigmat}) and (\ref{colort}) in
 eqs.~(\ref{corra})--(\ref{corrc}) and incorporating eqs.~(\ref{wfrca}) and
(\ref{wfrcb}),
 we get the one-loop matching conditions at $M_G$
\begin{eqnarray}
m^2_{H_{i}}(M_G)&=&m^2_{{\cal H}_i}(M_G)+
\frac{\lambda^2}{4\pi^2}(m^2_{{\cal H}_{1}}
+m^2_{{\cal H}_{2}}+m^2_{\Sigma}+A^2_{\lambda})
\left[\frac{3}{4}\ln\frac{M_{\Sigma_{3}}}{M_G}
+\frac{3}{20}\ln\frac{M_{\Sigma_{1}}}{M_G}
+\frac{3}{2}\ln\frac{M_{H_C}}{M_G}-\frac{6}{5}\right]\nonumber\\
&+&\frac{\lambda^2}{4\pi^2}\left[\frac{9}{10}
(m^2_{\Sigma}
+A_{\lambda}B_{\Sigma})+\frac{3}{4}
(m^2_{{\cal H}_{1}}+m^2_{{\cal H}_{2}}+2A_{\lambda}B_H)\right]\,
 ,\label{oneloopmata}\\
 m^2_i(M_G)&=&m^2_{10}(M_G)+
\frac{n_ih_t^2}{8\pi^2}(2m^2_{10}+m^2_{{\cal H}_{2}}+A^2_t)
\left[\ln\frac{M_{H_C}}{M_G}-\frac{1}{2}\right]\nonumber\\
&+&\frac{n_ih_t^2}{16\pi^2}
(m^2_{{\cal H}_{1}}+m^2_{{\cal H}_{2}}+2A_{t}B_H)\, ,\ \ \ \ \ \ n_i=1,2,3\
 {\rm for}\ i=U,Q,E\, .
\label{oneloopmatb}
\end{eqnarray}
Eq.~(\ref{oneloopmata}) corresponds to the one-loop
corrected eq.~(\ref{su5c}).
For $\mu_\Sigma\ll M_V\approx 2\times 10^{16}$ GeV, the logarithmic term of
eq.~(\ref{oneloopmata})
can lead to a large deviation from  eq.~(\ref{su5c}).
 For example, taking $M_G=M_V\sim
M_{H_C}\sim 10^3\mu_\Sigma$,
 $\lambda\approx 1$ and assuming $m^2_i = A^2_i = m^2_0$ at tree-level,
we have $m^2_{H_{i}}(M_G)\approx 0.6m^2_0$. The non-logarithmic terms of
eq.~(\ref{oneloopmata})
are smaller ($\sim 10\%$) and tend to cancel out for equal SSB parameters.
It is interesting to note that in the regions of the MSSM parameter space
 where the EWSB requires a high degree of fine tuning \cite{examples,penn594},
 a $10\%$ GUT correction to the soft Higgs masses can
destabilize the minimum.
The GUT threshold corrections to (\ref{su5a})
[given in eq.~(\ref{oneloopmatb})],
 are typically small since $M_{H_C}$ is forced to
be close to $M_V$ ($M_{H_C}\gtrsim M_V$ from proton decay and
$M_{H_C}\lesssim 2 M_V$ to stay in the perturbative regime \cite{proton}).
Nevertheless,
it is important to stress that, unlike other GUT effects,
such corrections contribute to the
 mass-splitting between light fields embedded in the same
 SU(5) representation.

Finally, the one-loop contribution to the trilinear term can be computed in
the same way, and is given by
\begin{eqnarray}
\Delta{A_{i}}& =& \frac{\lambda^2}{4\pi^2}A_{\lambda}
\left[\frac{3}{4}\ln\frac{M_{\Sigma_{3}}}{{\cal Q}}
+\frac{3}{20}\ln\frac{M_{\Sigma_{1}}}{{\cal Q}}
+\frac{3}{2}\ln\frac{M_{H_C}}{{\cal Q}}-\frac{6}{5}\right]\nonumber\\
&+&\frac{n_ih^2_t}{8\pi^2}A_t\left[\ln\frac{M_{H_C}}{\cal Q}
-\frac{1}{2}\right]\, ,
\label{lasteq}
\end{eqnarray}
where $n_i=3,2,3$ for $i=t,b,\tau$.



\begin{table}
\caption{Patterns of non-universality
in the minimal SU(5) model.
$\lambda \approx 0$
could lead to a too rapid proton decay
via dimension-five operators.
}
\label{table:t0}
\begin{tabular}{cccc}
$\tan\beta$ range & Pattern ($\lambda \approx 1$) &
Pattern ($\lambda \approx 0$) & choice\\
\hline
low ($1 - 2$) & $(i)$ & $(iii)$ & $(a)$\\
intermediate and large ($\gtrsim 50$)
& $(ii)$ & $(iv)$ & $(b)$ \\
\end{tabular}
\end{table}

\begin{table}
\caption{The low-energy spectrum is calculated for
$m_t^{pole} = 160$ GeV and $\tan\beta = 1.25$,
assuming universality at the listed scale.
We list, respectively,
the universal gaugino and scalar masses, the $D$-term parameter
and the trilinear parameter, along with the weak-scale predictions for
the $\mu$ parameter, the gluino, chargino and LSP masses,
the LSP eigenvector bino, wino and two Higgsino components,
the heavier neutralino masses, the light and heavy Higgs
boson masses, first and second family scalar quark
$\tilde{q}$ and lepton $\tilde{l}$ masses,
the $t$ and $b$-scalar masses and mixing $\tilde{t}_{1} =
-\sin\theta_{\tilde{t}}\tilde{t}_{R}
+ \cos\theta_{\tilde{t}}\tilde{t}_{L}$, and third family scalar lepton
masses.
$\lambda= 1$ (and $\lambda^{'} = 0.1$) at $M_{G}$. The $\sim$ implies
a rough average of the relevant masses.
Equal values of the model building [low-energy] parameters
eq. (7) [eq. (8)] are used in the first and second [first and third]
columns.
The first and second columns correspond to Fig. 1.
All masses are in GeV.}
\label{table:t1}
\begin{tabular}{c c c c }
scale & $M_{P}$ & $M_{G}$ & $M_{G}$ \\
$M_{1/2}$ & 165 & 165& 180\\
$m_{0}$& 987 & 987 & 991 \\
$M_{D}^{2}$ & 0 & 0& 0\\
$A_{0}/m_{0}$ & $-2$ & $-2$ & $-0.64$ \\
$\mu$& $-1755$ & $-2038$ & $-1988$ \\
$M_{\tilde{g}}$ & 499 & 460 & 499 \\
$m_{\chi^{+}_{1,\,2}}$ & 151, $-1758$ &138, $-2041$ & 150, $-1991$ \\
$m_{\chi^{0}_{1}}$ & 76 & 69 & 75 \\
$a_{11}$ & $-0.9988$ & $-0.9990$ & $-0.9990$ \\
$a_{12}$ & 0.0256 & 0.0243 & 0.0228 \\
$a_{13}$ & 0.0420 & 0.0368 & 0.0375 \\
$a_{14}$ & $-0.0055$ & $-0.0047$ & $-0.0048$ \\
$m_{\chi^{0}_{2}}$ & 151 & 138 & 150 \\
$m_{\chi^{0}_{3,\,4}}$ & 1755, $-1759$ & 2038, $-2041$ & 1988, $-1992$\\
$m_{h^{0}}$ & 75 & 75 & 79 \\
$m_{H^{0},\,A^{0},\, H^{+}}$ & $\sim 2435$ & $\sim 2945$ & $\sim 2895$ \\
$m_{\tilde{q}}$ & $\sim 1080$ & $\sim 1060 $& $\sim 1078$ \\
$m_{\tilde{l}}$ & $\sim 997$ & $\sim 992 $& $\sim 997$ \\
$m_{\tilde{t}_{1,\,2}}$ & 517, 811 & 250, 858& 395, 887 \\
$\sin\theta_{\tilde{t}}$ & 0.87 & 0.93 & 0.94 \\
$m_{\tilde{b}_{1,\,2}}$ & 735, 1075 & 794, 1059& 834, 1075 \\
$\sin\theta_{\tilde{b}}$ & 0.01 & 0.02 & 0.02 \\
$m_{\tilde{\tau}_{1,\,2},\,\tilde{\nu}_{\tau}}$ &729, $\sim 999$ &986, $\sim
995$
&993, $\sim 999$ \\
\end{tabular}
\end{table}

\begin{table}
\caption{Same as in Table II except it corresponds to Fig. 2.}
\label{table:t2}
\begin{tabular}{c c c c }
scale & $M_{P}$ & $M_{G}$ & $M_{G}$ \\
$M_{1/2}$ &400 &400 &436 \\
$m_{0}$&0  &0 &219 \\
$M_{D}^{2}$ &0 &0 &0 \\
$A_{0}/m_{0}$ &0 &0 &0.67 \\
$\mu$&$-1558$  &$-1348$ &$-1539$  \\
$M_{\tilde{g}}$ &1155  &1065  &1155  \\
$m_{\chi^{+}_{1,\,2}}$ &362, $-1562$ &333, $-1352$ &362, $-1542$  \\
$m_{\chi^{0}_{1}}$ &178  &163  &178  \\
$a_{11}$ & 0.9996 & 0.9995 & 0.9996 \\
$a_{12}$ & $-0.0099$ & $-0.0122$ & $-0.0100$ \\
$a_{13}$ & $-0.0256$ & $-0.0293$ & $-0.0258$ \\
$a_{14}$ & 0.0036 & 0.0042 & 0.0036 \\
$m_{\chi^{0}_{2}}$ & 362 & 333 &362  \\
$m_{\chi^{0}_{3, \,4}}$ &1558, $-1563$ &1348, $-1353$ &1539, $-1544$  \\
$m_{h^{0}}$ &64  & 62 & 63 \\
$m_{H^{0},\,A^{0},\, H^{+}}$ & $\sim 2097 $ & $\sim 1808 $ & $\sim 2081$ \\
$m_{\tilde{q}}$ & $1040 - 1090$ & $937 - 976 $& $1038 -1079$ \\
$m_{\tilde{l}}$ & $318 - 384$ & $157 - 290  $& $278 - 384$ \\
$m_{\tilde{t}_{1,\,2}}$ &822, 1001  &754, 907 &814, 993 \\
$\sin\theta_{\tilde{t}}$ &0.97  & 0.97 &0.98  \\
$m_{\tilde{b}_{1,\,2}}$ &982, 1037 &888, 937 &974, 1038 \\
$\sin\theta_{\tilde{b}}$ & 0.01 &0.01  &0.01  \\
$m_{\tilde{\tau}_{1,\,2},\,
\tilde{\nu}_{\tau}}$ &309, $\sim 382$ &157, $\sim 289$
&277, $\sim 382$ \\

\end{tabular}

\end{table}

\begin{table}
\caption{Same as in Table II except $\lambda(M_G) = 0.1$
and it corresponds to Fig. 3.}
\label{table:t3}
\begin{tabular}{c c c c }
scale & $M_{P}$ & $M_{G}$ & $M_{G}$ \\
$M_{1/2}$ &175 &175 &191  \\
$m_{0}$&375  &375 &387 \\
$M_{D}^{2}$ &0 &0 &0 \\
$A_{0}/m_{0}$ &1 &1 &0.73 \\
$\mu$&969  &946 &992  \\
$M_{\tilde{g}}$ &527  &486  &527  \\
$m_{\chi^{+}_{1, \,2}}$& 149, 977&136, 953  &149, 1000  \\
$m_{\chi^{0}_{1}}$ &77  &70  &77  \\
$a_{11}$ &$-0.7025$& $-0.7024$ & $-0.7028$ \\
$a_{12}$ &$-0.0718$& $-0.0726$ & $-0.0700$ \\
$a_{13}$ & 0.7068 & 0.7067 & 0.7068 \\
$a_{14}$ & 0.0423 & 0.0436 & 0.0411 \\
$m_{\chi^{0}_{2}}$ &149  &136  &149  \\
$m_{\chi^{0}_{3,\,4}}$ &$-969$, 979 &$-946$, 956 &$-992$, 1002 \\
$m_{h^{0}}$ &64  & 57 &66  \\
$m_{H^{0},\,A^{0},\, H^{+}}$ & $\sim 1363$ & $\sim 1327$ & $\sim 1392$ \\
$m_{\tilde{q}}$ & $\sim 610$ & $\sim 570 $& $\sim 607$ \\
$m_{\tilde{l}}$ & $\sim 405$ & $\sim 390 $& $\sim 405$ \\
$m_{\tilde{t}_{1,\,2}}$ &189, 631  &145, 613 &202, 644 \\
$\sin\theta_{\tilde{t}}$ &0.80  &0.81  &0.82  \\
$m_{\tilde{b}_{1,\,2}}$ &498, 601 &482, 565 &518, 601 \\
$\sin\theta_{\tilde{b}}$ &0.06  &0.07  &0.07  \\
$m_{\tilde{\tau}_{1,\,2},\,
\tilde{\nu}_{\tau}}$&340, $\sim 410$ &381, $\sim 395 $
&394, $\sim 410$ \\

\end{tabular}

\end{table}

\begin{table}
\caption{Same as in Table II except
$m_t^{pole} = 180$ GeV, $\tan\beta = 42$, and it corresponds to Fig. 4.
The last two columns list scenarios with non-vanising $D$-terms,
e.g., in SU(5)$\times$U(1). The latter two also provides a
crude approximation of the minimal SO(10) scenario.
Note that $M_{D}^{2} \neq 0$ splits $\tilde{q}$, $\tilde{l}$,
etc. according to their SU(5) embedding.
The value of $M_{D}^{2}$ which is used
in the last column
is the minimal
value still consistent with EWSB for the given set of parameters.}
\label{table:t4}
\begin{tabular}{c c c c c c}
scale & $M_{P}$ & $M_{G}$ & $M_{G}$ &$M_{P}$ & $M_{P}$ \\
$M_{1/2}$ &89 &89 &96 & 89& 89\\
$m_{0}$& 977 &977 &978 &977&977\\
$M_{D}^{2}$ &0 &0 &0&$ +0.16m_{0}^{2}$&$-0.05m_{0}^{2}$ \\
$A_{0}/m_{0}$ &0 &0 & 0.07&0&0 \\
$\mu$& $-645$ &$-214$ &$-225$&$-849$&$-563$  \\
$M_{\tilde{g}}$ & 280 &259  &280 &280&280 \\
$m_{\chi^{+}_{1,\,2}}$ &80, $-646$  &74, $-215$  &80, $-226$
&80, $-849$&80, $-563$ \\
$m_{\chi^{0}_{1}}$ &40  &35  &38&40&40  \\
$a_{11}$ & 0.9976 & 0.9709 & 0.9739&0.9986&0.9968 \\
$a_{12}$ & 0.0107 & 0.1408 & 0.1279&0.0049&0.0153 \\
$a_{13}$ & $-0.0469$ & $-0.1221$ & $-0.1162$&$-0.0362$&$-0.0532$ \\
$a_{14}$ & 0.0504 & 0.1504 & 0.1452&0.0379&0.0582 \\
$m_{\chi^{0}_{2}}$ &79  &65  &71 &79&78 \\
$m_{\chi^{0}_{3, \,4}}$ &$-651$, 652 &$-230$, 239 &$-240$, 249
&$-854$, 854&$-569$, 571 \\
$m_{h^{0}}$ & 114 &113  &114 &114&114 \\
$m_{H^{0},\,A^{0},\, H^{+}}$ & $\sim 492 $ & $\sim 567 $ & $\sim 568$&
 $\sim 920$ & $\sim 220$ \\
$m_{\tilde{q}}$ & $\sim 1007 $ & $\sim 1000  $& $\sim 1007$ &
$\sim 1080,\, 744$ &$\sim 980,\,1077$ \\
$m_{\tilde{l}}$ & $\sim 979 $ & $\sim 977  $& $\sim 979$ &
$\sim 715,\, 1055$ &$\sim 1055,\,954$ \\
$m_{\tilde{t}_{1,\,2}}$ & 691, 790  & 598, 731 &601 ,737& 795, 881& 653, 757 \\
$\sin\theta_{\tilde{t}}$ &0.96  &0.98  &0.97& 0.97&0.96  \\
$m_{\tilde{b}_{1,\,2}}$ &743, 869  & 708, 814 &712 ,818&482, 875&725, 939 \\
$\sin\theta_{\tilde{b}}$ &0.44  & 0.16 & 0.17 &0.98&0.20\\
$m_{\tilde{\tau}_{1,\,2},\,
\tilde{\nu}_{\tau}}$ &818, $\sim910 $ &803, $\sim 869$
&804, $\sim 897$&$\sim 592$, 919&793, $\sim 985$ \\

\end{tabular}

\end{table}


\begin{figure}
\caption{
Same as in Fig. 6 except for the light CP-even Higgs boson mass
$m_{h^{0}}$ prediction.
}
\label{fig:fig8}
\end{figure}

\begin{figure}
\caption{
Same as in Fig. 7 except for the light CP-even Higgs boson mass
$m_{h^{0}}$ prediction.
}
\label{fig:fig9}
\end{figure}

\begin{figure}
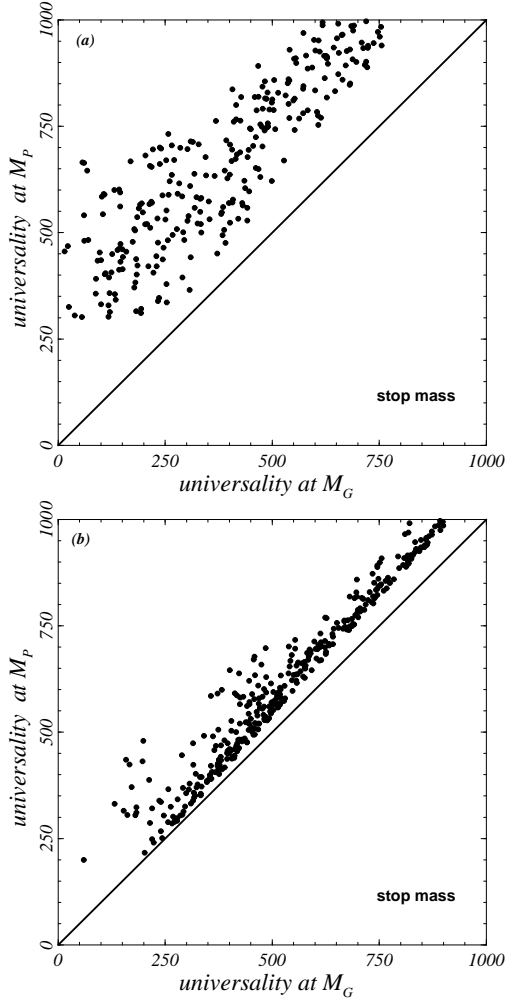

\caption{
Same as Fig. 5 except the prediction for the light $t$-scalar mass
$m_{\tilde{t}_{1}}$.
}
\label{fig:fig10}
\end{figure}

\begin{figure}
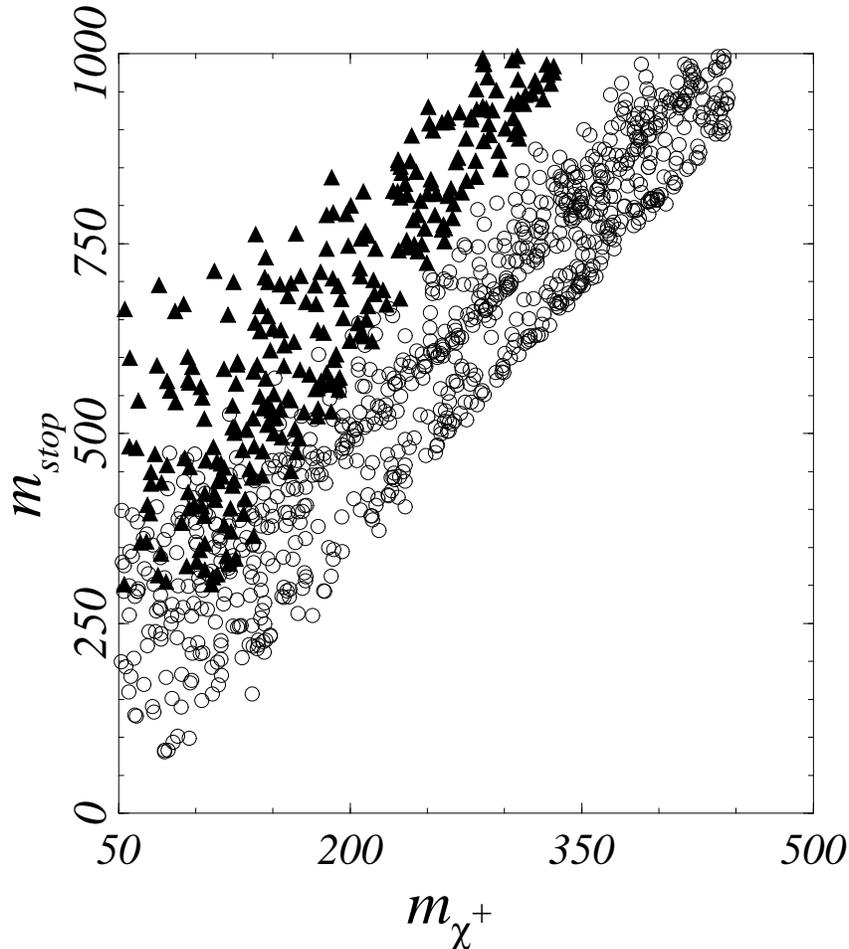

\caption{
Scatter plot of the light chargino $\chi^{+}_{1}$
$vs.$ the light $t$-scalar $\tilde{t}_{1}$
masses (in GeV)
within the allowed parameter space (see above) and for
$m_t^{pole} = 160$ GeV and $\tan\beta = 1.25$.
Filled triangles [circles] correspond to universality
at $M_{P}$ [$M_{G}$] and
$\lambda(M_{G}) = 1$, $\lambda^{'}(M_{G}) = 0.1$.
}
\label{fig:fig11}
\end{figure}

\begin{figure}
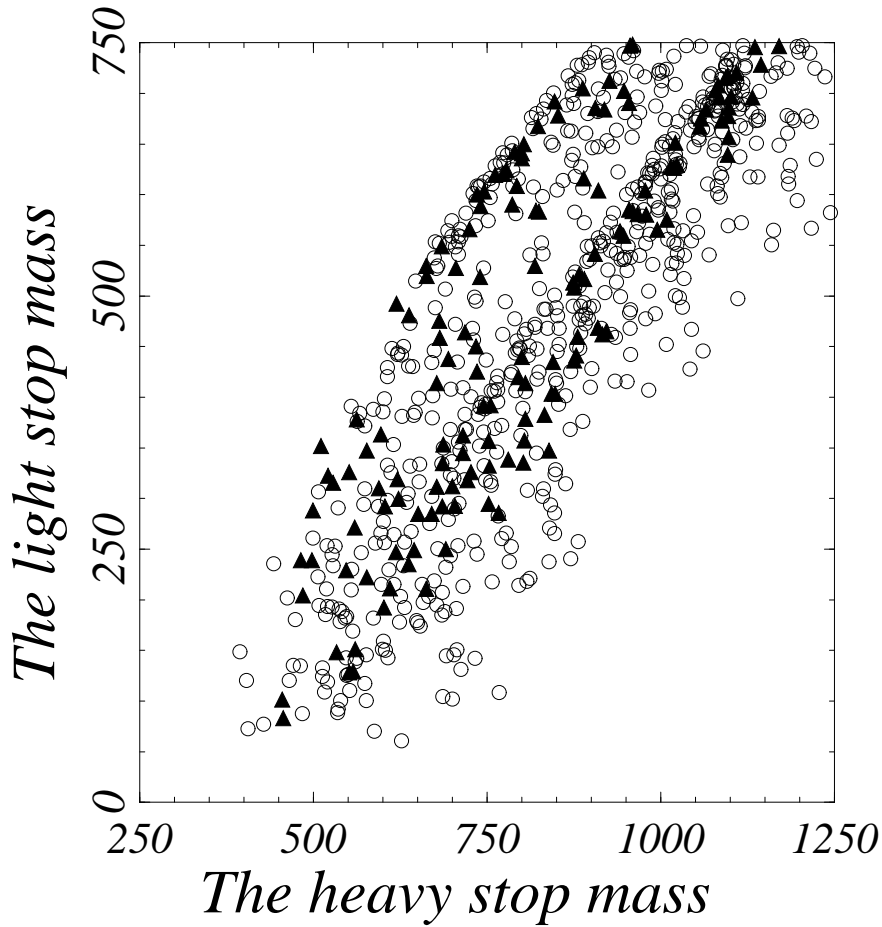

\caption{
Same as in Fig. 11 except
the light ${\tilde{t}_{1}}$ $vs.$ heavy ${\tilde{t}_{2}}$
masses.
}
\label{fig:fig12}
\end{figure}

\begin{figure}
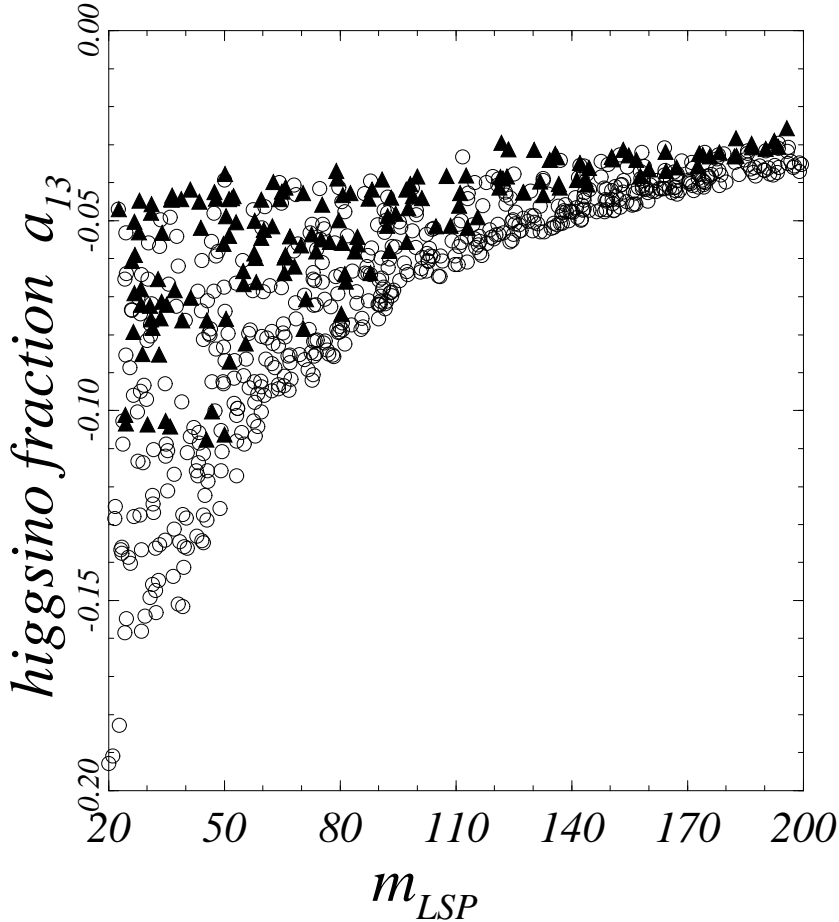

\caption{
Same as in Fig. 11 except the LSP mass $m_{\chi_{1}^{0}}$ (in GeV)
$vs.$ its Higgsino fraction $a_{13}$ and
for $m_t^{pole} = 180$ GeV and $\tan\beta = 42$.
}
\label{fig:fig13}
\end{figure}


\end{document}